\documentclass[12pt]{article}

\pdfoutput=1
\usepackage{mathptmx}
\usepackage{amsmath,amssymb,amstext,amsthm,amscd,mathrsfs,eucal,bm,slashed}
\usepackage{graphicx,color}
\usepackage{array}
\usepackage{cite}
\usepackage{tikz}
\usetikzlibrary{shapes.geometric}
\usetikzlibrary{decorations.pathmorphing}
\usetikzlibrary{arrows,shapes,positioning}
\usetikzlibrary{decorations.markings}
\tikzstyle arrowstyle=[scale=1]
\tikzstyle directed=[postaction={decorate,decoration={markings,
    mark=at position .65 with {\arrow[arrowstyle]{stealth}}}}]
\tikzstyle reverse directed=[postaction={decorate,decoration={markings,
    mark=at position .65 with {\arrowreversed[arrowstyle]{stealth};}}}]

\usepackage{hyperref}
\hypersetup{
  pdftitle   = {Quantum fields in curved spacetime},
  pdfkeywords = {},
  pdfauthor  = {S. Hollands, R.~M.~Wald},
  pdfcreator = {\LaTeX\ with package \flqq hyperref\frqq}
}

\newcommand{\half}{\tfrac12}
\newcommand{\cB}{\mathcal{B}}

\newcommand{\cC}{\mathcal{C}}

\newcommand{\rI}{{\rm I}}

\newcommand{\TT}{\mathbb{T}}
\newcommand{\RR}{\mathbb{R}}
\newcommand{\CC}{\mathbb{C}}

\newcommand{\ZZ}{\mathbb{Z}}

\newcommand{\eA}{\mathscr{A}}
\newcommand{\eB}{\mathscr{B}}

\newcommand{\eD}{\mathscr{D}}

\newcommand{\eG}{{\mathscr G}}
\newcommand{\eH}{\mathscr{H}}

\newcommand{\eL}{\mathscr{L}}
\newcommand{\eM}{\mathscr{M}}

\newcommand{\eQ}{\mathscr{Q}}

\newcommand{\eW}{\mathscr{W}}

\renewcommand{\Im}{\mathrm{Im}}

\newcommand{\dv}{{\rm dV}}

\renewcommand{\half}{\tfrac{1}{2}}

\newcommand{\Q}{{\rm Q}}

\newcommand{\D}{{\rm D}}

\renewcommand{\H}{\mathcal{H}}
\newcommand{\M}{\mathscr{M}}
\newcommand{\myid}{{\bf 1}}
\renewcommand{\pounds}{{\mathscr L}}

\newcommand{\mr}{\mathbb{R}}
\newcommand{\mc}{\mathbb{C}}
\renewcommand{\O}{\Phi}

\newcommand{\T}{{\rm T}}
\newcommand{\e}{{\rm e}}

\renewcommand{\P}{{\bf P}}

\DeclareMathOperator{\WF}{WF}

\renewcommand{\k}{{\bf k}}

\newcommand{\rf}[1]{[\![#1]\!]}

\theoremstyle{plain}
\newtheorem{lemma}{Lemma}

\newtheorem{theorem}[lemma]{Theorem}

\theoremstyle{definition}

\def\ben{\begin{equation}}
\def\een{\end{equation}}
\def\bena{\begin{eqnarray}}
\def\eena{\end{eqnarray}}
\def\non{\nonumber}
\newcommand{\MUNCH}[1]{\relax}

\allowdisplaybreaks
\begin{document}
%


\title{Quantum fields in curved spacetime}



\footnotetext[1]{Universit\"{a}t Leipzig, Institut f\"{u}r Theoretische Physik, Br\"{u}derstrasse 16, D-04103 Leipzig, FRG}
\footnotetext[2]{Enrico Fermi Institute and Department of Physics, University of Chicago, Chicago, IL 60637, USA}


\maketitle
\begin{abstract}

We review the theory of quantum fields propagating in an arbitrary, classical, globally hyperbolic spacetime. Our review emphasizes the conceptual issues arising in the formulation of the theory and presents known results in a mathematically precise way. Particular attention is paid to the distributional nature of quantum fields, to their local and covariant character, and to microlocal spectrum conditions satisfied by physically reasonable states. We review the Unruh and Hawking effects for free fields, as well as the behavior of free fields in deSitter spacetime and FLRW spacetimes with an exponential phase of expansion. We review how nonlinear observables of a free field, such as the stress-energy tensor, are defined, as well as time-ordered-products. The ``renormalization ambiguities'' involved in the definition of time-ordered products are fully characterized. Interacting fields are then perturbatively constructed. Our main focus is on the theory of a scalar field, but a brief discussion of gauge fields is included. We conclude with a brief discussion of a possible approach towards a nonperturbative formulation of quantum field theory in curved spacetime and some remarks on the formulation of quantum gravity.

\end{abstract}

\section{The nature of Quantum Field Theory in Curved Spacetime}

Quantum field theory in curved spacetime (QFTCS) is the theory of quantum fields propagating in a background,
classical, curved spacetime $(\M, g)$. On account of its classical treatment of the metric, QFTCS cannot be a fundamental theory of nature. However, QFTCS is expected to provide an accurate description of quantum phenomena in a regime where the effects of curved spacetime may be significant, but effects of quantum gravity itself may be neglected. In particular, it is expected that QFTCS should be applicable to the description of quantum phenomena occurring  in the early universe and near (and inside of) black holes---provided that one does not attempt to describe phenomena occurring so near to singularities that curvatures reach Planckian scales and the quantum nature of the spacetime metric would have to be taken into account.

It should be possible to derive QFTCS by taking a suitable limit of a more fundamental theory wherein the spacetime metric is treated in accord with the principles of quantum theory. However, this has not been done---except in formal and/or heuristic ways---simply because no present quantum theory of gravity has been developed to the point where such a well defined limit can be taken. Rather, the framework of QFTCS that we shall describe in this review has been obtained by suitably merging basic principles of classical general relativity with the basic principles of quantum field theory in Minkowski spacetime. As we shall explain further below, the basic principles of classical general relativity are relatively easy to identify and adhere to, but it is far less clear what to identify as the ``basic principles'' of quantum field theory in Minkowski spacetime. Indeed, many of the concepts normally viewed as fundamental to quantum field theory in Minkowski spacetime, such as Poincare invariance, do not even make sense in the context of curved spacetime, and therefore cannot be considered as ``fundamental'' from the viewpoint of QFTCS. By forcing one to re-think basic concepts, such as the notions of ``vacuum state'' and ``particles,'' QFTCS has led to deep insights into the nature of quantum field theory---and one may hope that it will provide significant guidance towards the development of quantum gravity itself.

The fundamental ideas upon which classical general relativity is based are that (i) all aspects of spacetime structure are described by the  topological and differential (i.e., manifold) properties of events together with a Lorentz signature metric $g$, and (ii) the metric and matter fields\footnote{One of the truly remarkable aspects of general relativity is that no new ``matter field'' need be introduced to describe gravitation, i.e., all physical phenomena normally attributed to ``gravity'' are, in fact, described by $g$.} are dynamical; furthermore their evolution is locally determined. More precisely, the metric and the tensor (and/or spinor) fields describing matter satisfy partial differential equations---namely, Einstein's equation together with the equations of motion for the matter fields---that have a well posed initial value formulation, so that these fields are uniquely determined (up to ``gauge'') from their initial data within a suitable domain of dependence. In particular, in classical general relativity, there is no non-dynamical, ``background structure'' in the laws of physics apart from the manifold structure of events. This lack of background structure in classical general relativity is usually referred to as the ``covariance'' or ``coordinate invariance'' of the theory; a ``preferred set of coordinates'' defined independently of the metric would provide non-dynamical, background structure.

It is much more difficult to identify the fundamental ideas upon which quantum field theory in Minkowski spacetime is based. One can attempt to formulate the quantum theory of a field in Minkowski spacetime by decomposing the field into modes and applying the rules of quantum mechanics to each mode. For a free field, each mode is an independent harmonic oscillator and one can obtain a mathematically sensible quantum field theory in this manner, although even here one encounters infinite expressions for quantities that are nonlinear in the fields. A well known example of this general phenomenon is that one obtains an infinite expression for the total energy (and energy density) of the field, as can be seen by adding the zero-point energies and/or energy densities of the infinite number of modes. The situation is considerably worse for interacting (i.e., nonlinear) fields, wherein one immediately encounters ill defined and/or infinite expressions in the calculation of essentially all physical quantities, arising from the fact that modes of arbitrarily high energies seemingly contribute to low energy processes. Historically, it it appears to have been generally assumed in its earliest days that the quantum field theory description of nature would break down at, say, the energy scale of elementary particles, and there was no reason to presume that it was a mathematically consistent theory. However, starting from the early 1950s, it was gradually understood how to give mathematically consistent rules to produce well defined expressions for physical quantities to all orders in perturbation theory for renormalizable theories such as quantum electrodynamics. This process culminated in the works of Bogliubov, Parasiuk
and of Hepp and Zimmermann, with important practical improvements (dimensional regularization) being given later by `t~Hooft and Veltmann. It was also seen that the predictions of quantum field theory give truly excellent agreement with experiment---as they have continued to do through the present, LHC era. In the 1950s and 1960s, major progress was made toward putting quantum field on a mathematically sound footing via the development of the axiomatic~\cite{sw}, algebraic~\cite{haag}, and constructive~\cite{glimm} approaches. Nevertheless, the prevailing attitude toward quantum field theory today is not very different from what it was in its earliest days, namely, that it is not a fundamental theory but merely a valid description of quantum field modes up to some cutoff in energy (now assumed to be at a much higher energy scale than would have been assumed in its earliest days). At the present time, relatively little attention is generally paid to the issue of whether quantum field theory can be given a mathematically precise and consistent formulation---as compared with such issues as the ``fine tuning'' that would be necessary to give small values to the cosmological constant and Higgs mass if one views quantum field theory as the quantum theory of the modes of fields lying below some energy cutoff.

Our view is that it is very important to determine if quantum field theory can be given a mathematically precise and consistent formulation as a theory in its own right---and to provide such a formulation if it can be given. This is not because we believe that quantum field theory should be a ``final'' theory of nature; indeed, we do not believe that a quantum theory of the spacetime metric can be formulated within the existing framework of quantum field theory. However, even if quantum field theory has only a limited domain of validity, it is important to understand precisely what questions are well posed within its framework and how the answers to these questions are to be obtained. In this way, the predictions of quantum field theory can be made with clarity and precision, and hints may be provided for some of the features that might be expected to survive in a more fundamental theory that supersedes quantum field theory.

What are the ``basic principles'' of quantum field theory in Minkowski spacetime? The observables of the theory are the tensor fields representing the fundamental constituents of matter, together with ``composite fields,'' such as the stress-energy tensor, derived from these matter fields. A key basic principle of quantum field theory is that each observable field, $\O(x)$, at each spacetime point $x$ should be represented as an operator. These operators will satisfy nontrivial algebraic relations, such as commutation relations. However, there are two important caveats to this statement that $\O(x)$ is represented as an operator.

The first caveat is that, even in the case of a free Klein-Gordon scalar field $\phi$---where, as already mentioned above, a quantum field theory can be formulated by ordinary quantization of the independent modes---it can be seen that one cannot make mathematical sense of $\phi(x)$ as an operator at a sharply defined point $x$, since modes of arbitrarily high frequency and short wavelength contribute to $\phi(x)$. However, $\phi(x)$ does make sense as a distribution, i.e., by ``averaging'' $\phi(x)$ with a smooth function of compact support, $f(x)$, one effectively eliminates the arbitrarily high frequency and short wavelength oscillations and thereby obtains a well defined expression for the quantum field. Thus, quantum fields are operator-valued distributions. The distributional nature of quantum fields is the source of most, if not all, of the mathematical difficulties arising in quantum field theory. Nonlinear operations involving distributions are intrinsically ill defined, and one will typically
get infinite answers if one attempts to evaluate nonlinear functions of a distribution via mode expansions or other procedures.

The second caveat is that the word ``operator'' presumes that there is some unique underlying Hilbert space of states on which this operator will act. However, even for a free field, there are an infinite number of unitarily inequivalent representations of the fundamental commutation relations. (This contrasts sharply with the situation for a quantum mechanical system with a finite number of degrees of freedom, where the Stone-von Neuman theorem asserts that, under mild additional assumptions, all such representations are unitarily equivalent.) In Minkowski spacetime, a preferred representation normally can be chosen based upon the additional requirement that the representation contain a Poincare invariant state (``the vacuum''). However, no criterion analogous to this can be applied in a general curved spacetime. As discussed further at the end of this section, it would therefore seem much more natural to view the algebraic relations satisfied by the field observables---rather than the choice of representation---to be fundamental. Thus, in quantum field theory, we will take as a basic principle that {\em the quantum fields $\O(x)$ are distributions valued in an algebra.}

Another basic principle of quantum field theory in Minkowski spacetime that the fields should ``transform covariantly'' under Poincare transformations. The Poincare group is the isometry group of the metric, $\eta$, of Minkowski spacetime, but a general, curved spacetime will not admit any isometries. Nevertheless, ``Poincare invariance'' may be viewed as a special relativistic version of the above general relativistic requirement of ``covariance,'' i.e., that quantum field theory in curved spacetime be constructed out of the classical spacetime metric $g$ and the fundamental quantum fields $\phi$, without any additional ``background structure.'' Furthermore, this construction should be local in nature. We will take as a basic principle that {\em quantum field theory should be locally and covariantly constructed.} We will give precise meaning to this statement in section~\ref{sec3}.

An additional basic principle of quantum field theory in Minkowski spacetime is the requirement of positivity of energy. Since the notion an energy operator for a quantum field theory in Minkowski spacetime is normally defined in terms of the transformation properties of the field under time translations, this requirement cannot be straightforwardly generalized to QFTCS. Nevertheless, one can formulate local conditions on the quantum field theory---known as microlocal spectral conditions---that correspond to positivity of energy in Minkowski spacetime and make sense in curved spacetime. We will take as a basic principle that {\em the quantum field theory should satisfy suitable microlocal spectral conditions.} We will give precise meaning to this statement in Appendix~\ref{appA}.

Finally, an additional principle of quantum field theory in MInkowski spacetime that is usually taken to be fundamental is the existence of a unique, Poincare invariant state. However, this condition has no analog in a general curved spacetime as a condition on existence or uniqueness of states\footnote{As argued in \cite{hw5}, the existence of an operator product expansion may be viewed as a generalization of this condition to curved spacetime; see subsection~\ref{sec:ope} below.}, and we will not attempt to impose any condition of this nature.

Thus, we seek a formulation of quantum field theory in curved spacetime that implements the three basic principles written in italics above. In the remainder of this section, we briefly describe some of the standard approaches that have been used to formulate quantum field theory in Minkowski spacetime and explain why they do not appear suitable for the formulation of QFTCS. We will then describe the approach that we shall adopt.

Many discussions of quantum field theory are based upon a notion of ``particles,'' and focus almost entirely on the calculation of the $S$-matrix, describing the scattering of particles. For a free field in Minkowski spacetime, a notion of ``vacuum state'' and ``particles'' can be defined in a natural and precise manner. If an interacting field behaves like a free field in the asymptotic past and future, one can define asymptotic particle states. The $S$-matrix provides the relationship between the ``in'' and ``out'' particle descriptions of the states, and thereby directly yields the dynamical information about the interacting field that is most relevant to laboratory experiments in high energy physics. However, the use of an $S$-matrix description as a fundamental ingredient in the formulation of QFTCS is unsuitable for the following reasons. Although natural notions of ``vacuum state'' and ``particles'' can be defined for a free field in stationary spacetimes, no such natural notions exist in a general curved spacetime. The difficulty is not that a notion of ``particles'' cannot be defined at all in a general curved spacetime but rather that many notions exist and none appears preferred. Although it may be possible (and useful) to define an $S$-matrix in spacetimes that become asymptotically stationary in a suitable manner in the past and future, many of the spacetimes of greatest interest in QFTCS are cosmological spacetimes or spacetimes describing gravitational collapse, where singularities occur in the asymptotic past and/or future. If one wishes to apply QFTCS to such spacetimes, it clearly would be preferable to formulate it in a manner that does not require one to define a notion of ``particles'' near singularities before one can even pose a well defined question. Furthermore, even if the spacetime of interest is suitably asymptotically stationary in the past and future, many of the most interesting physical questions are concerned with the local dynamical behavior of the fields at finite times rather than the particle-like description of states at asymptotically early and late times. For example, one may wish to know the expected stress-energy tensor of a quantum field in order to estimate the ``back reaction'' effects of the quantum field on the dynamics of the spacetime. An $S$-matrix would not be useful for such a calculation.

In many discussions of quantum field theory in Minkowski spacetime, Euclidean methods play an important role in both the formulation of the theory and in calculational techniques. Minkowski spacetime can be viewed as a real $4$-dimensional section of a complex $4$-dimensional manifold with complex metric, which contains another $4$-real-dimensional section (of ``imaginary time'') on which the metric is positive definite. If one can define a quantum field theory in a suitable manner on this ``Euclidean section,'' a quantum field theory on Minkowski spacetime can then be obtained via analytic continuation. Since it is much easier to make sense of formal expressions in the Euclidean setting than in the Lorentzian setting, Euclidean methods have been employed in most of the attempts to rigorously define interacting quantum field theories and in most of the methods employed to regularize and renormalize quantities in perturbative quantum field theory~\cite{glimm}. Euclidean methods can be generalized so as to apply to static, curved spacetimes, where the transformation ``$t \to it$'' takes one from a static Lorentzian spacetime to a Riemannian space. However, a general curved spacetime will not be a real section of a complex manifold that also contains a real section on which the metric is Riemannian. Thus, although it should be possible to define ``Euclidean quantum field theory'' on curved Riemannian spaces~\cite{komu}, there is no obvious way to connect such a theory with quantum field theory on Lorentzian spacetimes. Thus, if one's goal is to define quantum field theory on general Lorentzian spacetimes, it does not appear fruitful to attempt to formulate the theory via a Euclidean approach.

Finally, by far the most prevalent approach taken towards the formulation of quantum field theory in Minkowski spacetime is to write down a formal functional integral expression for an effective action. Suitable functional derivatives of this expression are then interpreted as providing the correlation functions of the quantum field in its vacuum state. Thus, one will have defined the quantum field theory if one can make sense of this functional integral and its functional derivatives. The difficulty with using a functional integral approach to formulate QFTCS is that, in effect, it requires one to single out a preferred state in order to define the theory---namely, the state for which the correlation functions are being given. This is not a difficulty in Minkowski spacetime, where Poincare invariance naturally selects a preferred state and, furthermore, Euclidean methods are available to make sense of the functional integral for this preferred state. However, as previously indicated above, no analogous notion of a preferred state exists in a general curved spacetime without symmetries. As in the above discussion of $S$-matrix approaches to the formulation of QFTCS, we do not believe that it will be fruitful to formulate QFTCS via an approach that requires one to define a preferred state in order to define the theory\footnote{Note that this objection does not apply to the formulation of quantum field theory in (complete) Riemannian spaces, where one has a unique Green's functions for Laplacian-like operators, and, thus, a ``preferred state" for a free field. However, as discussed above, there does not appear to be any way of relating quantum field theory in Riemannian spaces to quantum field theory on Lorentzian spacetimes.}.

For the above reasons, we shall adopt the ``algebraic viewpoint'' for the formulation of QFTCS. The basic idea of this approach is to take the relations satisfied by the quantum fields---such as commutation relations and field equations---as the fundamental starting point of the theory. To define the theory, one must specify the complete set of algebraic relations satisfied by the fundamental field and composite fields. As we shall see, this can be done for a free field. In addition, we can naturally define time ordered products, enabling one to give a perturbative construction of interacting quantum field theory. However, it is far less clear as to how to define appropriate algebraic relations so as to give a non-perturbative definition of interacting quantum field theory; we will return to this issue in subsection~\ref{sec:ope}. Once the algebraic relations have been given, states are defined to be positive linear functions on the algebra of quantum fields. The GNS construction then shows that every state in this sense arises as a vector in a Hilbert space that carries a representation of the field algebra, thus connecting the algebraic notion of states with usual notions of states in quantum theory. The key point is that one can formulate QFTCS via the algebraic approach in a manner that does not require one to single out a preferred state in order to define the theory.

We begin in the next section by formulating QFTCS for a free scalar field, taking into account only the fundamental field observables. We also discuss some key applications, including the Hawking and Unruh effects and quantum effects arising from inflation. In section~\ref{sec3}, we show how a wide class of nonlinear field observables can be defined for a free scalar field. We then describe the perturbative construction of QFTCS for interacting fields. We also discuss the construction of QFTCS for gauge fields. Finally, in section~\ref{sec5} we discuss ideas that may lead towards a nonperturbative formulation of interacting QFTCS and we make some remarks on the formulation of quantum gravity from the perspective of QFTCS.

\section{Free quantum fields}
\label{sec2}

In this section, we provide a precise formulation of the theory of a free Klein-Gordon field in curved spacetime, insofar as the fundamental field observable, $\phi$, is concerned. (Observables that are nonlinear in $\phi$---i.e., ``composite fields''---will be introduced in the next section.) We then discuss some key applications of the theory, namely, the Unruh effect, quantum field theory in deSitter spacetime, the Hawking effect, and cosmological perturbations. 

\subsection{Formulation of linear QFTCS via the algebraic approach (without
nonlinear observables).}
We now describe how to define the quantum field theory of a real, linear
Klein-Gordon field $\phi$ on a $d$-dimensional, curved, Lorentzian
spacetime $(\M,g)$ along the lines sketched in the previous section.
We begin with the classical Klein-Gordon field, which satisfies
\ben\label{KG}
(\square_g - m^2)\phi = 0
\een
where $\square_g = g^{\mu\nu} \nabla_\mu \nabla_\nu$ is the D'Alembertian operator associated with $g$. In order that~\eqref{KG} have a well-posed initial value
formulation, we restrict consideration---here and throughout this article---to globally hyperbolic spacetimes. By definition, a globally hyperbolic spacetime is a time oriented spacetime that possesses
a ``Cauchy surface,'' $\Sigma$, i.e., a smoothly
embedded $(d-1)$-dimensional spacelike submanifold with the property that if
$\gamma: \mr \to \M$ is any inextendible causal curve, then $\gamma$
intersects $\Sigma$ precisely once.~\footnote{It is a theorem~\cite{sanchez} that any globally
hyperbolic spacetime has the topology of a direct product,
$\M \cong \Sigma \times \mr$.}
The classical Klein-Gordon equation with source $j$
\ben
(\square_g - m^2)\phi = j
\label{kgs}
\een
(where $j$ is an arbitrary, fixed, smooth function on spacetime) has a well posed initial value formulation on globally hyperbolic spacetimes in the following sense.
Let $n$ denote the unit normal to $\Sigma$. Then, given any pair $(f_0,f_1)$ of smooth functions
on $\Sigma$, there exists a unique
solution $\phi$ to the Klein-Gordon equation~\eqref{kgs} such that
\ben\label{data}
\phi |_{\Sigma} = f_0, \quad n^\mu \nabla_\mu \phi |_{\Sigma} = f_1 \, .
\een
Furthermore, solutions to the initial value problem have a causal dependence upon
the initial data and the source in the sense that if $x \in J^+(\Sigma)$,
then the solution $\phi$ to the above initial value problem will not change at $x$ if
we change the initial data $(f_0,f_1)$ outside $J^-(x) \cap \Sigma$,
or if we change the source outside of $J^-(x) \cap J^+(\Sigma)$. Here
\ben
J^\pm(S) \equiv \{x \in \M \mid \text{$\exists$ causal future/past directed
curve from $y \in S$ to $x$}\},
\een
denotes the ``causal future/past'' of a set $S \subset
\M$. Similarly, if $x \in J^-(\Sigma)$, then $\phi(x)$ will not change if
we change the initial data $(f_0,f_1)$ outside $J^+(x) \cap \Sigma$,
or if we change the source outside of $J^+(x) \cap J^-(\Sigma)$. Finally, the solution $\phi$
depends continuously on $(f_0, f_1)$ and $j$ in a suitable sense.

We can define the retarded and advanced
propagators $E^\pm$ of the Klein-Gordon equation as follows. For $j \in C_0^\infty(\M)$,
the advanced solution $E^- j$ is the unique solution to the Klein-Gordon equation
with source $j$ such that its initial data vanish on some arbitrary
Cauchy surface $\Sigma$ such that $J^-(\Sigma)$ contains ${\rm supp} j$,
with the opposite definition for the retarded solution $E^+ j$.
The propagators may be viewed as maps
\ben
E^\pm: C^\infty_0(\M) \to C^\infty(\M) \, ,
\een
or alternatively as distributional kernels on $\M \times \M$.
As distributions, $E^\pm$ satisfy the differential equation
\ben\label{propeq}
(\square_g - m^2)E^\pm (x,y) = \delta(x,y) \ ,
\een
where the Klein-Gordon operator acts on the first variable, $x$, in
the sense of distributions.
The support properties are
\ben
{\rm supp} \, E^\pm \subset \{(x,y) \in \M \times \M \mid
y \in J^\pm(x) \}.
\een
The advanced and
retarded propagators are related by exchanging $x$ with $y$.
The anti-symmetric combination
\ben\label{Edef}
E = E^+ - E^-
\een
is called ``commutator function''\footnote{Further discussion concerning the construction and local expansion of $E^\pm$ can e.g.
be found in~\cite{bar}.}.

As described in the Introduction, we will formulate QFTCS for a Klein-Gordon field
by defining a suitable algebra, $\eA(\M,g)$ of quantum observables. In this section, we will consider
only the algebra of observables generated by the fundamental field $\phi$. An enlarged algebra that includes ``composite fields'' will be defined in the next section.
The construction of $\eA(\M,g)$ will take into account: 1)
the distributional nature of the field $\phi(x)$, 2) the field equation, 3)
the real character of $\phi$ and
and 4) the symplectic structure of the classical phase space of this
theory. We construct $\eA(\M,g)$ by starting with the free *-algebra
generated by a unit $\myid$ and
elements $\phi(f)$, with $f \in C^\infty_0(\M)$, and factoring by the following relations:
\begin{enumerate}
\item
{\bf Linearity:} $\phi(c_1f_1 + c_2 f_2) = c_1 \phi(f_1) + c_2
\phi(f_2)$ for all $c_1,c_2 \in \mc$.
\item
{\bf Field equation:} $\phi((\square_g - m^2)f) = 0$,
\item
{\bf Hermitian field:} $\phi(f)^* = \phi(\bar f)$,
\item
{\bf Commutator:} $[\phi(f_1), \phi(f_2)] = iE(f_1,f_2) \myid$.
\end{enumerate}
Item 1) incorporates the distributional character of the field; informally
we write
\ben
\phi(f) = \int_\M \phi(x) f(x) \ dv_g
\een
and we think of $\phi(x)$ as an $\eA$-valued distribution.
We refer to the algebra $\eA$ as ``abstract,'' because no
reference has been made to any representation. In fact, we are not
supposed to think, a priori, of the elements of $\eA$ as
operators on a particular Hilbert space, in the same way as an
abstract Lie-algebra is defined irrespective of a particular
representation. 2) incorporates the field equation in a distributional sense.
3) expresses that the field is real, and 4) implements the usual quantum mechanical relationship
between classical Poisson-brackets and commutators. It also incorporates
Einstein causality, because $E(x,y) = 0$ if $x,y$ are spacelike related.

Rather than working with fields $\phi(f)$ that are smeared with test functions
$f$ on $\M$, we can also equivalently view $\phi$ as being ``symplectically smeared'' with solutions, $F$, having initial
data of compact support on some Cauchy surface $\Sigma$. The correspondence
is the following. If $f \in C^\infty_0(\M)$, then $F=Ef$ is a source free solution having
initial data~\eqref{data} of compact support. Conversely, given a solution $F$ with initial
data of compact support, there exists a test function $f$---unique up to addition of $(\square_g - m^2) h$ for $h \in C^\infty_0$---such
that $F = Ef$. Defining $\phi[F] = \phi(f)$ under this correspondence between $F$ and $f$, we can then informally write the
field as
\ben\label{KGip}
\phi[F] = \int_\Sigma (F \nabla_\mu \phi - (\nabla_\mu F) \phi )n^\mu dS \ .
\een
We can think of $\pi = n^\mu \nabla_\mu \phi |_\Sigma, \varphi = \phi |_\Sigma$
as canonically conjugate variables and the commutation relation (4) then
corresponds to $[\pi(x), \varphi(y)] = i\delta_\Sigma(x,y) \myid$;
for details, see, e.g.~\cite{dimock} or lemma 3.2.1 of \cite{wald2}.

A {\bf physical state}, $\omega$, is simply an ``expectation
value functional,'' i.e., a linear map $\omega: \eA(\M,g) \to \mc$ satisfying the
normalization condition $\omega(\myid) = 1$, and positivity,
$\omega(a^* a) \ge 0$ for all $a \in \eA$. Any state is, by construction,
specified by the collection $(W_n)_{n \ge 1}$ of its ``$n$-point functions'',
\ben
W_n(f_1, \dots, f_n) \equiv \omega(\phi(f_1) \cdots \phi(f_n)) \ .
\een
The KG-equation, condition 1), is translated into the fact that
$W_n$ is a distributional solution in each entry. Condition 2)
implies that $W_n$ is an $n$-times multilinear functional on
$C^\infty_0(\M)$, which one normally requires to be distributional
(i.e. continuous in the appropriate sense). The commutator
condition 4) is translated into a linear condition which in the
simplest case $n=2$ is
\ben
W_2(x,y) - W_2(y,x) = i E(x,y) \ .
\label{commu}
\een
Positivity is translated into a rather complicated hierarchy of
conditions on the multi-linear functionals $W_n$, the simplest of which is
\ben
W_2(\bar f, f) \ge 0 \qquad \text{for all $f \in C_0^\infty$.}
\label{w2pos}
\een

Given two states $\omega, \omega'$,
one can form a new state by forming any convex linear combination
$\lambda \omega + (1-\lambda) \omega'$, where $0 \le \lambda \le 1$.
A state which cannot be written as a non-trivial convex linear combination
of others is called {\bf pure}.

The notion of algebraic state is in principle sufficient to answer all
physical questions about the field observables. In particular, the specification of a state
$\omega$ directly yields the expected values of all powers of $\phi(f)$ for all, say real, test functions $f$.
It follows from the classical ``Hamburger moment problem'' (see e.g.~\cite{simon}) that
there is a unique\footnote{A classic result about the Hamburger moment problem is that 
measure $d\nu$ is {\em unique} if the moments $m_n$ satisfy the growth condition $$\sum_n 1/\sqrt[2n]{m_{2n}}  = \infty,$$
see e.g.~\cite{akhizer}. This criterion can be applied straightforwardly to the free KG field $\phi(f)$, but not to non-linear observables such as Wick powers $\phi^k(f), k>2$ discussed in sec.~\ref{nonlinear}, where $m_{2n} = O((kn)!)$. Thus, for general, nonlinear observebles, the determination of the probability distribution from the state remains open.}  probablity measure $d\nu(\lambda)$ such that
\ben
\int_\RR \lambda^n d\nu(\lambda) = \omega(\underbrace{\phi(f) ... \phi(f)}_{n}) \equiv m_n \qquad
\text{for all $n\ge 1$.}
\een
The probability that
an observation of $\phi(f)$ will yield a value within $[a,b]$ when the field is in state $\omega$
is then given by
\ben
P_{a,b} = \int_{[a,b]} \lambda d\nu(\lambda) \ .
\een

The relationship between states in the algebraic sense as defined above and the usual notion of states as
vectors in a Hilbert space can be seen as follows:
First, assume that we have a representation $\pi: \eA(\M,g) \to \eH$ on a Hilbert space
with dense invariant domain $\eD \subset \eH$. If $\Psi$ is a non-zero vector in this domain,
then
\ben
\omega_\Psi(a) = \frac{(\Psi, \pi(a) \Psi)}{(\Psi,\Psi)}
\een
defines a state in the algebraic sense. More generally, any
sufficiently regular (with respect to $\eD$) density matrix on $\eH$---i.e., a non-negative, trace-class operator
$\rho \in {\mathscr I}_1(\eH)$---defines
an algebraic state $\omega_\rho$. Conversely, given any algebraic state $\omega$, there is a simple
construction---known as the {\bf GNS-construction}, or {\bf Wightman reconstruction argument}---that yields a Hilbert space $\eH$, a representation $\pi$ of $\eA$ on $\eH$ with invariant
domain $\eD \subset \eH$, and a vector $\Omega \in \eD$ such that the
algebraic state corresponding to $\Omega$ is $\omega$. As we shall explain further below, for the case of a Gaussian state, the GNS construction yields a Fock representation of $\eA$, with $\Omega$ being the vacuum vector of this Fock space.

The above correspondences show that the algebraic and Hilbert space formulations
of QFTCS are essentially equivalent. However, there is one important difference: There are many
unitarily inequivalent representations of the field algebra $\eA$. (Two representations are said to be unitarily
equivalent if there is an isometry $U: \eH \to \eH'$ such that $U \pi(a) U^* = \pi'(a)$ for
all $a \in \eA$.) The Hilbert space formulation requires one to
chose a ``preferred representation'' at the outset, while the algebraic formulation does not. For this reason, we feel that the formulation of QFTCS via the algebraic approach is conceptually superior.

The algebra $\eA$ admits states with rather
pathological properties of their $n$-point functions $W_n$. However, there is a natural criterion
to select physically reasonable states
that is motivated
from a variety of closely related considerations, including that (i) the ultra-high-frequency modes of the field should be essentially in their ground state, (ii) the short distance singular structure of the $n$-point functions $W_n$ should be similar to that of the $n$-point functions of the vacuum state in Minkowski spacetime, and (iii) the singular structure of the $W_n$'s
should be of ``positive frequency type''. A mathematically precise implementation of these requirements
in a general curved spacetime can be formulated in the language of wave front sets discussed in
Appendix~\ref{appA}, as first proposed in~\cite{rad1,rad2} and~\cite{bf1}. The criterion is that the $2$-point function have a wavefront set of the form
\ben\label{hadacon}
\WF(W_2) \subset \{ (x_1,k_1;x_2,k_2) \in T^*\M ^2 \setminus 0 \mid k_1 \in \dot V^+, k_2 \in \dot V^-, k_1 \sim -k_2 \} \ ,
\een
and that each ``connected $n$-point function'' $\omega^c_n(\phi(x_1),...,\phi(x_n))$ for $n\neq 2$ (defined in eq.~\eqref{cnpf} below) is smooth\footnote{
The last requirement may actually be shown to be a consequence of the first condition~\cite{sanders1}.}.
Here $V^+ \subset T^* \M$ is the collection of all non-zero, future directed
time-like or null co-vectors (and similarly $V^-$), and the relation $\sim$ holds between two covectors if they are tangent to a null geodesic 
in $(\M,g)$ and are parallel transported into
each other . Such states are called ``{\bf Hadamard states}''. As we will see in the next section, it will be necessary to restrict to Hadamard states in order to extend the action of the state
to an enlarged algebra of field observables that includes all polynomials in the field and its derivatives at the same spacetime point. In particular, it will be necessary to restrict to Hadamard states when
considering the perturbative expansion of
interacting QFTCS.

One can prove~\cite{rad1,rad2}
that the above definition of a Hadamard state
is equivalent to the following local condition~\cite{kaywald} on the 2-point function together with the requirement that, globally, there be no singularities at spacelike separations:
For every convex,
normal neighborhood $U \subset \M$, the two-point function has the form
(in $d=4$ dimensions; similar expressions hold in arbitrary $d$):
\bena\label{localhada}
W_2 &=& \frac{1}{4\pi^2} \left[
\frac{u}{\sigma + i0^+ t} + \left( \sum_{n=0}^N v_n
  \sigma^n \right) \log(\sigma
+ i0^+ t)
\right] \nonumber \\
&+& \quad \text{some $N$-times continuously differentiable function $R_{N,\omega}$}\\
&\equiv& H_N + R_{N,\omega} \, . \nonumber
\eena
Here, the spacetime arguments are understood as $(x,y) \in U
\times U$, and we mean that the above formula should
hold for every $N$,
with (different) remainders $R_{N,\omega} \in C^{N}(U \times U)$.
$\sigma$ is the signed squared geodesic distance, and $t=T(x)-T(y)$
for some (in fact, any) global time function $T: \M \to \RR$.
The functions $u, v_n$ are determined by certain local ``transport equations''~\cite{dewitt} and
also appear in similar local forms of the advanced
and retarded Green's function for the operator $\square_g
-m^2$. In particular, $H_N$
is locally and covariantly defined in terms of the metric, and
is hence the same for {\it any} Hadamard state. By contrast,
the remainder $R_{N,\omega}$ depends on the state.

The Hadamard condition does not single out a particular state, but a {\em class} of states.
Existence of
a large class of Hadamard state on {\em any} globally hyperbolic spacetime
can be established by a deformation
argument~\cite{fulling} combined with microlocal techniques, or by methods from
the theory of pseudo-differential operators~\cite{junker,gerard}.

We now discuss two important classes of states: Gaussian states and thermal states.

\vspace{.5cm}
{\bf Gaussian States:} Gaussian states (also called ``quasi-free states'')
are defined by the condition that the ``connected $n$-point functions''
$\omega^c_n(\phi(f_1),...,\phi(f_n))$ vanish for all $n > 2$, where
$\omega^c_n: \eA \times ... \times \eA \to \CC$ is defined by\footnote{
Here the exponentials are to be understood in the
sense of a formal series.}
\ben
\omega^c_n(a_1,\dots,a_n) \equiv \frac{\partial^n}{\partial t_1 \dots \partial t_n} \log \bigg\{ \omega\bigg(
\e^{t_1 a_1} \cdots \e^{t_n a_n} \bigg) \bigg\} \bigg|_{t_i=0} \ , \quad a_i \in \eA \ .
\label{cnpf}
\een
 Thus, the $n$-point functions of Gaussian states can be expressed
entirely in terms of their 1- and 2-point functions. For a Gaussian state, positivity will hold if and only
if~\eqref{w2pos} is satisfied. Thus, there exist a wide class of Gaussian states.

Any Gaussian state, $\omega$, can be expressed as the ``vacuum state'' in a Fock representation
of $\eA$. To see this explicitly, let $W_2$ be the 2-point function of $\omega$. On the complex linear
space $C^\infty_0(\M, \CC)$ of smooth complex-valued functions on $\M$ of compact support, define the
inner product $\langle f | h \rangle = W_2(\overline f, h)$. This
is hermitian and positive, $\langle f | f \rangle \ge 0$, but contains
degenerate vectors, such as elements of the form $f = (\square_g -
m^2)h$. Let $\frak h$ be the factor space of $C^\infty_0(\M,\CC)$, divided by the
degenerate vectors. The elements in this space can be
identified more concretely with a subspace of complex valued smooth solutions to the KG-equation, corresponding to ``positive frequency modes'', see below.
The completion of $\frak h$, denoted by the same symbol, is a Hilbert space, usually referred to
as the ``1-particle
space''. Let $\eH$ be the bosonic Fock space over $\frak h$,
\ben\label{fockspaceh}
\H = \CC \oplus \bigoplus_{n \ge 1} \underbrace{({\frak h} \otimes_S \cdots \otimes_S {\frak h})}_{n}\, ,
\een
where $\otimes_S$ is the symmetrized tensor product.
A representation, $\pi$, of $\eA$ on $\eH$ can then be defined by
\ben\label{phihbrack}
\pi[\phi(f)] = a([f])^\dagger + a([f])
\een
where $a([f])$ is the annihilation operator associated with the
equivalence class of $f$ in $\frak h$. The vacuum vector $\Omega$ given
by the element $|1,0,0,\dots \rangle$ in Fock-space then corresponds to $\omega$, and,
as already mentioned above, $(\eH, \pi, \Omega)$ is precisely the GNS triple arising from the GNS construction.

A closely related construction often used in practice to construct Gaussian pure states is the following.
Suppose that we have a set of smooth, complex-valued ``mode functions'' $u_\xi(x)$ that are
solutions to eq.~\eqref{KG}, and which are labelled by $\xi \in X$ in some
measure space $(X, d\mu)$. We assume that for each $f \in C^\infty_0(\M,\RR)$,
the map
\ben
X \owns \xi \mapsto Kf(\xi) \equiv \int_\M \overline{u_\xi(x)} f(x) \ dv_g \in \CC
\een
is in $L^2(X,d\mu)$, and in fact that the (real linear)
span of such vectors is dense in $L^2(X,d\mu)$. We also assume that the mode functions are such that
\ben\label{wronsk}
\Im \langle Kf_1 | Kf_2 \rangle_{L^2(X,d\mu)} = \half E(f_1, f_2) \qquad
\text{for all $f_1, f_2 \in C_0^\infty(\M)$.}
\een
(These properties are equivalent to the statement that the collection of modes
$(u_\xi)_{\xi \in X}$ is ``complete in the KG-norm''.). Then, clearly
\ben\label{modes}
W_2(f_1, f_2) = \langle Kf_1 | Kf_2 \rangle_{L^2(X,d\mu)} \quad
\Rightarrow \quad W_2(x,y) = \int_X d\mu(\xi)  u_\xi(x) \overline{u_\xi(y)}
\een
defines the 2-point function of a Gaussian state, which can be shown to be pure.
Its GNS-representation is thus constructed as above. $K$ is clearly well-defined on the equivalence classes $[f]$, and it can also be shown that
it provides a bounded isomorphism $K: {\frak h} \to L^2(X,d\mu)$. Hence, in this case,
we may consider $\eH$ as the bosonic Fock-space over $L^2(X,d\mu)$, and we may
informally write the representative of the field on this Fock space as
\ben
\pi(\phi(x)) = \int_X d\mu(\xi) [u_\xi(x) a_\xi^{} + \overline{u_\xi(x)} a_\xi^\dagger]
\een
This is the usual form of the field in the vacuum representation
on Minkowski space, where $\xi = {\bf k} \in \RR^3 = X$, $d\mu({\bf k}) = d^3 {\bf k} / (2\pi)^3$, and $u_{\bf k}({\bf x},t)
= 
e^{-i\omega_{\bf k} t + i{\bf kx}}/\sqrt{2\omega_{\bf k}}$. In this example, the mode functions have positive frequency
with respect to the global time translations, and similar constructions  are available also on other 
spacetimes with a globally defined time-like isometry. 

Given two pure Gaussian states $\omega, \tilde \omega$ with 2-point functions $W_2$ respectively
$\tilde W_2$, one may ask when their associated GNS-representations are
{\bf unitarily equivalent}. A necessary condition is that there be a constant
$c$ such that
\ben
c^{-1} W_2(f, f) \le \tilde W_2(f,f) \le c \, W_2(f, f)
\een
for all $f \in C^\infty_0(\M,\RR)$. Since it is easy to construct states violating this condition, one sees that
there is in general a large class of inequivalent representations. If the condition is
satisfied, it follows that there is a bounded linear
operator $S$ on ${\frak h}$ such that $\langle [f] | [f]
\rangle_{\tilde {\frak h}}
= \langle [f] | S [f] \rangle_{{\frak h}}$. A sufficient criterion
which ensures that the two states give rise to unitarily equivalent representations is that
\ben\label{crit}
{\rm tr} [(1-S)(1-S^\dagger)]^{\half} < \infty \, .
\een
A sufficient and also necessary criterion\footnote{The case of non-Gaussian
states is open.} can be found e.g. in~\cite{araki}. The operator $1-S$ in essence
characterizes the difference between the 2-point functions. It is therefore plausible that
if their difference is smooth -- as happens e.g. if both states
are  Hadamard -- and the manifold $\M$ has compact slices,
then~\eqref{crit} should hold, and the representations should be unitarily
equivalent. This can indeed be shown\footnote{
For Gaussian states that are not pure the generalization of unitary 
equivalence is ``quasi-equivalence''~\cite{araki}. Ref.~\cite{verch}
actually deals with this more general situation.
}~\cite{verch}. On the other hand, for the case of a non-compact
Cauchy surface, the representations can be unitarily
inequivalent if the 2-point functions have sufficiently different long-range behavior.

\vspace{.5cm}
{\bf Thermal States:} If the spacetime $(\M, g)$ has a complete time-like Killing vector field $\xi$,
one can define the notion of a thermal state relative to the time evolution
generated by this Killing vector field. There is an elegant version of this
notion referred to as the ``KMS-condition,'' which can be formulated directly
in terms of the expectation value functional $\omega$, without making
reference to any Hilbert-space representation.
The KMS condition is formulated as follows:
Let $\varphi_t$ be
1-parameter group of isometries $\varphi_t: \M \to \M$ generated by a Killing vector field $\xi$. We
define an action $\alpha_t: \eA \to \eA$ of our 1-parameter family of
isometries on the algebra $\eA$ of fields on $(\M,g)$ by setting
$\alpha_t(\phi(f_1) \cdots \phi(f_n)) = \phi(f^t_1) \cdots
\phi(f^t_n)$, where $f^t(x) = f(\varphi_{-t}(x))$. Since the retarded
and advanced fundamental solutions, and the field equation, are invariant under
$\varphi_t$, it follows that $\alpha_t$ respects the algebraic
relations in $\eA$, i.e., it is an automorphism. In fact, from the
composition law for the isometries $\varphi_t$ it immediately follows
that $\alpha_t \circ \alpha_s = \alpha_{t+s}$.
In this situation, a state $\omega$ is called a {\bf KMS-state} at inverse temperature $\beta$
with respect to $\alpha_t$ if the following two conditions are
satisfied:  
\begin{enumerate}
\item For any collection of $a_i \in
  \eA$, the function $\underline t=(t_1, \dots, t_n) \mapsto F_{a_1, ..., a_n}(\underline t)$  defined by
\ben
 F_{a_1, ..., a_n}(\underline t) = \omega(\alpha_{t_1}(a_1) \cdots \alpha_{t_n}(a_n))
\een
has an analytic continuation to the strip
\ben
{\frak T}^\beta_n = \{ (z_1, \dots, z_n) \in \CC^n \mid 0 < \Im(z_j) - \Im(z_i) < \beta \ , \ \
1 \le i < j \le n \} \ ,
\een
This function is required to be bounded and continuous at the boundary.
\item On the boundary, we have
\ben\label{kmscond}
\begin{split}
&F_{a_1, ..., a_n}(t_1, \dots, t_{k-1}, t_k + i\beta, \dots, t_n+i\beta) \\
=&F_{a_k, ..., a_n, a_1, ... , a_{k-1}}(t_k, \dots, t_n, t_1, \dots, t_{k-1}) \ .
\end{split}
\een
\end{enumerate}
Note that the definition of a KMS-state only assumes an algebra $\eA$
and the existence of a 1-parameter family of automorphisms. The notion
of a KMS-state is therefore not tied to the particular example $\eA =
\eA(\M,g)$ and the particular 1-parameter group of automorphisms
$\alpha_t$ considered here. It is thus a definition of a very general
nature, applicable to many quantum systems, see e.g.~\cite{bratelli} for further discussion. 
In the case of $C^*$-algebras (algebras of
bounded operators),
the condition for $n=2$ implies the remaining ones, but this is not generally
the case for the case of unbounded operator algebras considered here. It is
however the case for the concrete algebra $\eA$ considered here
if we restrict attention to Gaussian states. In this case, the condition $n=2$
also implies that the state is Hadmard~\cite{sahlmann}. 

Let us now motivate the above technical definition by explaining
its relation to the usual notion of thermal
equilibrium state in statistical mechanics. Consider a self-adjoint Hamiltonian $H$
defined on a Hilbert space with spectrum bounded from below, and
suppose that $Z_\beta = {\rm tr} \ \e^{-\beta H} < \infty$ (which cannot hold 
unless the spectrum of $H$ is discrete). The
standard definition of a Gibbs state is
$\omega ( a ) = {\rm tr} (a \e^{-\beta H})/Z_\beta$,
 where $a$ is e.g. any (say) bounded
operator on $\eH$. Let $\alpha_t(a)$ be defined in this example by
$\alpha_t(a)=\e^{itH} a \e^{-itH}$, i.e., it
describes
the usual time evolution of an observables $a$
in ordinary quantum mechanics. Then, using
that the spectrum of $H$ is
bounded below, we easily see that,
\ben
F_{a,b}(z) =
Z_\beta^{-1} {\rm tr} (a \e^{iz H} b \e^{-iz H}\e^{-\beta H} )
\een
is holomorphic in the strip $0 < \Im(z) < \beta$, because in this range $\e^{-\beta H}$ 
 provides a sufficient ``damping''  to make the trace finite. Furthermore, using
the cyclicity of the trace, we have
\bena
F_{a,b}(t+is) &=&
Z_\beta^{-1} \
{\rm tr} (\e^{-\beta H} a \e^{(it-s) H} b \e^{(-it+s)
  \hat H} ) \\
&=&
Z_\beta^{-1} \
{\rm tr} (\e^{-\beta H} \e^{(it+\beta-s)) H} b
\e^{(-it-\beta+s)
  H} a) \, .
\eena
From the first line we see
that  $F_{a,b}(t+is) \to
\omega(a \alpha_t(b))$ for $s\to 0^+$,
while we see from the second line that $F_{a,b}(t+is) \to
\omega(\alpha_t(b) a)$ for $s\to \beta^-$. Thus, (1) and (2) hold (for $n=2$), and 
therefore a Gibbs state in the usual sense is a
KMS-state in the sense of the above definition. The
idea behind the definition of a KMS-state is to turn this statement
around and define thermal equilibrium states by conditions 1) and 2).

A key technical advantage of the definition of a KMS-state is that it
still makes sense when a density matrix no longer exists, as usually
happens when the Cauchy surface $\Sigma$ is non-compact. The standard example of
this is Minkowski space (`thermodynamic limit'). In the standard GNS-representation $(\eH, \pi, \Omega)$ corresponding
to the vacuum state (described e.g. by the mode functions given above), the Hamiltonian
does not have the property that $\e^{-\beta H}$ is a trace-class operator on $\eH$, i.e.
no density matrix exists. Nevertheless, a Gaussian KMS-state can easily be defined in
terms of its 2-point function, given by
\ben\label{KMS0}
W_2(x_1,x_2) = \omega(\phi(x_1)\phi(x_2)) =
\frac{1}{(2\pi)^3} \int D^\beta_+(t_1-t_2-i0^+,{\bf p}) \, \e^{i{\bf p}({\bf x}_1-{\bf x}_2)} \, d^3 {\bf p} \ ,
\een
where Minkowski points are labelled by $x = (t, {\bf x}) \in \RR^{1,3}$, and where $D^\beta_+$ is given by
\ben\label{Ddef}
D^\beta_+(t,{\bf p}) = \frac{1}{2 \omega_{\bf p}} \frac{\cosh(\half \beta \omega_{\bf p} - i\omega_{\bf p}t)}{\sinh(\half \beta \omega_{\bf p})} \ ,
\qquad \omega_{\bf p} = \sqrt{{\bf p}^2 + m^2} \ .
\een
The 1-parameter family of isometries is simply given by $\varphi_T(t,{\bf x}) = (t+T,{\bf x})$, i.e. time translations.
The verification of the KMS-condition boils down to the condition on the 2-point function, which in
turn boils down to showing that the $z=(t+is)$-dependent distribution
$F_{x,y}(t+is) = W_2(x, \varphi_{t+is}(y))$ in $x,y \in \RR^{1,3}$ has distributional boundary
values $W_2(x,\varphi_t(y))$ resp. $W_2(\varphi_t(y), x)$ for $s \to 0^+$ resp. $s \to \beta^-$. This is
in turn directly seen to be a consequence of the functional relation
\ben
\lim_{s \to \beta^-} D^\beta_+(t-is,{\bf p}) = \lim_{s \to 0^+} D^\beta_+(-t+is,{\bf p})  \ .
\een

\subsection{Applications: Unruh effect, deSitter space, Hawking effect, inflationary perturbations}

We now discuss some concrete examples in order to illustrate the abstract ideas just given and to
present some of the important applications of QFTCS.

\vspace{.5cm}
{\bf a) Unruh effect:} A relatively simple and yet very important application of QFTCS
arises if we consider a ``wedge'' $W$ of Minkowski
spacetime and view it as a spacetime in its own right. Namely, let
\ben
W = \{x \in \RR^{1,3} \mid x_1 > |x_0| \},
\een
and
let $W$ be equipped with the Minkowski metric. Of course, this is not a curved spacetime, but
it is a globally hyperbolic spacetime that differs in essential ways from
Minkowski spacetime, e.g., all of its timelike and null geodesics are incomplete. The spacetime $(W, g)$ is called ``Rindler spacetime''.

Writing $U = -x_0 + x_1$ and $V = x_0 + x_1$, the metric of Rindler spacetime is
\ben
\begin{split}
g &= dU dV + dx^2_2 + dx^2_3  \\
&= \e^{a(u+v)} du dv + dx^2_2 + dx^2_3
\end{split}
\een
where $u$ and $v$ are defined by $U=\e^{au}$ and $V=\e^{av}$. Further introducing $(\eta, \xi)$ by
\ben
u = \xi - \eta, \quad v = \eta + \xi,
\een
the metric takes the form
\ben
g = \e^{2a\xi}(- d\eta^2 + d\xi^2) + dx^2_2 + dx^2_3 .
\een
The coordinates $\eta$ and $\xi$ are related to
the original global inertial coordinates $(x_0, x_1, x_2, x_3)$ of Minkowski spacetime by
\ben
\label{boosts}
\begin{split}
x_0 &= a^{-1} \e^{a\xi} \sinh a\eta\\
x_1 &= a^{-1} \e^{a\xi} \cosh a\eta.
\end{split}
\een

It is not difficult to see that the hypersurfaces, $\Sigma_\eta$, of constant $\eta$ are Cauchy surfaces for Rindler spacetime for all $\eta \in \RR$. Furthermore, for any $t \in \RR$, the transformation $\varphi_t: \eta \to \eta +
t$ is an isometry of the Rindler spacetime, which corresponds to Lorentz boosts of Minkowski spacetime. Indeed,
the key fact about Rindler spacetime is that the orbits of the Lorentz boosts are everywhere timelike and are complete in Rindler spacetime. Thus, Rindler spacetime is a static, globally hyperbolic spacetime, where the notion of ``time translations'' is defined by Lorentz boosts. Note that each Lorentz boost orbit in Rindler spacetime corresponds to the worldline of a uniformly accelerating observer in
Minkowski spacetime. The Lorentz boost orbits become null on the boundary of Rindler spacetime and, indeed, the hypersurfaces $U=0$ and $V=0$ of Minkowski spacetime comprise a bifurcate Killing horizon of the Lorentz boost Killing field, with surface gravity\footnote{Note that unlike in the analogous example of Schwarzschild (see below), there is in this case 
no canonical normalization of the Killing field.} $\kappa = a$.

As on every globally hyperbolic spacetime, we quantize the field
$\phi$ by viewing it (after smearing with a test function) as an
element of the associated abstract algebra $\eA(W,g)$ defined by the
relations 1) to 4) above. Actually, in the present context, these
relations are identical with those of the whole Minkowski spacetime,
because the advanced and retarded propagators, and hence $E$, are
locally the same. However, there is a difference in that the algebra
$\eA(W, g)$ only contains smeared elements of the form $\phi(f)$ for
test functions $f$ that are compactly supported in $W$ (and in particular, away from $\partial W$). Thus,
$\eA(W,g)$ may be viewed as a proper subalgebra of the algebra
associated with the entire Minkowski spacetime. Thus, we can obtain
a state on $\eA(W,g)$ by restricting the usual vacuum state on Minkowski spacetime, and view it as a state on
Rindler space. The 2-point function of this Gaussian state is (taking $m^2=0$ for simplicity)
\ben
W_2(x,y) = \frac{1}{2\pi^2 (x-y-i0^+ e)^2} \ ,
\label{2ptfun}
\een
where $e$ is any fixed future directed timelike vector, and the distributional boundary
value prescription is understood.

Since $W_2(x,y) \neq 0$ for any
$x \in W$ and $y \in W'$ (where $W'$ denotes the ``opposite wedge'' $x_1 < - |x_0|$),
it follows that there are correlations between field observables in $W$ and $W'$ and
that restriction of the
Minkowski vacuum to Rindler spacetime cannot yield a pure state. A key result is the following theorem,
which is a special case of the Bisognano-Wichmann theorem~\cite{haag} of axiomatic quantum field theory in Minkowski
spacetime:

\medskip

\begin{theorem}
The restriction of the Minkowski vacuum state to the Rindler algebra $\eA(W, g)$ is
a KMS-state with respect to the 1-parameter group of isometries given
by $\eta \to \eta + t$. The inverse temperature of this KMS-state is given
by
\ben
\beta = \frac{2\pi}{a} \, .
\label{beta}
\een
\end{theorem}
 To prove this claim, one has to verify the KMS-condition. This can be done in
a completely straightforward manner. For a Gaussian state such as
as that given here, it suffices to verify the KMS-condition for $a_1=\phi(f_1), a_2=\phi(f_2)$
with $f_1,f_2$ having their support inside $W$,
which in turn boils down to showing that the distribution
$F_{x,y}(t+is) = W_2(x, \varphi_{t+is}(y))$ in $x,y \in W$ has distributional boundary
values $W_2(x,\varphi_t(y))$ resp. $W_2(\varphi_t(y), x)$ for $s \to 0^+$ resp. $s \to \beta^-$.
This is an elementary computation done by transforming the 2-point function
into Rindler coordinates.

The Minkowski vacuum state is, of course, regular on the Rindler horizon and is invariant under Lorentz boosts, and it is the only Hadamard state on all of Minkowski spacetime that is Lorentz boost invariant. In fact, the uniqueness and KMS property (but not necessarily existence) of an isometry invariant Hadamard state can be proven to hold on any globally hyperbolic spacetime with a bifurcate Killing horizon \cite{kaywald}. Important further examples will be provided in the next two subsections.

The above theorem has an important physical interpretation, known as the {\bf Unruh effect}: If the field is in the Minkowski vacuum state, a uniformly accelerating observer in Minkowski spacetime---who may also be viewed as a static observer in Rindler spacetime---will ``feel himself'' immersed in a ``thermal bath of particles'' at inverse temperature \eqref{beta}. This can be explicitly seen by introducing a model ``particle detector'' and showing that it
will be suitably excited as a result of its interaction with the quantum field~\cite{unruh1}; for a recent treatment see e.g.~\cite{bievre}. This provides an excellent illustration of
why notions of ``vacuum state'' and ``particles'' cannot be considered to be fundamental in the formulation of
QFTCS. If the field is in the Minkowski vacuum state, an inertial observer will naturally declare that no ``particles'' are present,
whereas the accelerating observer will naturally declare that the Rindler wedge is filled with a thermal bath of particles.
However, there is no actual disagreement between these observers: They both agree that the field is in a Gaussian
state with two-point function \eqref{2ptfun}, and they will be in complete agreement on the probabilities for
measuring any field obserables.

\vspace{.5cm}
{\bf b) deSitter spacetime:}
Four dimensional (global) deSitter space $dS_4$ is the 4-dimensional hyperboloid
defined by the equation $Y \in \RR^5, Y \cdot Y = H^{-2}$ where $H >0$ is the
Hubble constant and the dot ``$\cdot$'' denotes the 5-dimensional Minkowskian
inner product with signature $(-++++)$. The metric is that induced
from the ambient space. $d$-dimensional deSitter space is defined in the same way.

\begin{figure}
\begin{center}
\begin{tikzpicture}[scale=.85, transform shape]
\fill[blue!25!,opacity=.3] (-4,2) rectangle (0,-2);
\draw (0,0) -- (-2,2) -- node[black,below, sloped]{$Z=-1$} (-4,0) -- node[black,above, sloped]{$Z=-1$} (-2,-2) -- node[black,above, sloped]{$Z=1$} (0,0) node[right]{$x$};
\draw (-2.7,0) node[black,right]{$|Z|<1$};
\draw (3,0) node[black,left]{$|Z|<1$};
\draw (-2,-2) -- (-4,-2) --  (-4,2) -- (-2,2);
\draw (0,0) -- (2,2) -- (4,0) -- (2,-2) -- node[black,above, sloped]{$Z=1$}(0,0);
\draw[blue] (0,2) -- (0,-2);
\draw (0,0) -- node[black,below, sloped]{$Z=1$}(-2,2) --
node[below,black]{$Z>1$} node[above,black]{${\mathscr I}^+$} (2,2) -- node[black,below, sloped]{$Z=1$} (0,0);
\draw (0,0) -- node[black,above,sloped]{$Z=1$} (-2,-2) -- node[black,above]{$Z>1$} node[below,black]{${\mathscr I}^-$}(2,-2)  --  (0,0);
\draw[->, very thick] (-2,3) node[above left] {$J^+(x) =$ future of $x$} -- (-.8,1.5);
\draw[->, very thick] (-5,0) node[left] {$Z<-1$} -- (-3.5,1);
\draw[->, very thick] (-5,0) -- (-3.5,-1);
\draw (3,0) node[black,left]{$|Z|<1$};
\draw (2,-2) -- (4,-2) -- node[black,right]{} (4,2) -- (2,2);
\end{tikzpicture}
\end{center}
\caption{
\label{dS1}
Conformal diagram and values of the point-pair invariant $Z=Z(x,y)$ as $y$ is varied and $x$ is kept fixed.
For the sake of easier visualization, we are giving the diagram in the case of $d=2$ dimensional deSitter spacetime, where
the left and right vertical boundaries are to be identified.
For $d>2$ dimensions, the diagram would basically consist of only the \textcolor{blue}{shaded} ``left half'', with the vertical boundary lines corresponding 
to the north- and south pole of the $S^{d-1}$ Cauchy surface.
}
\end{figure}
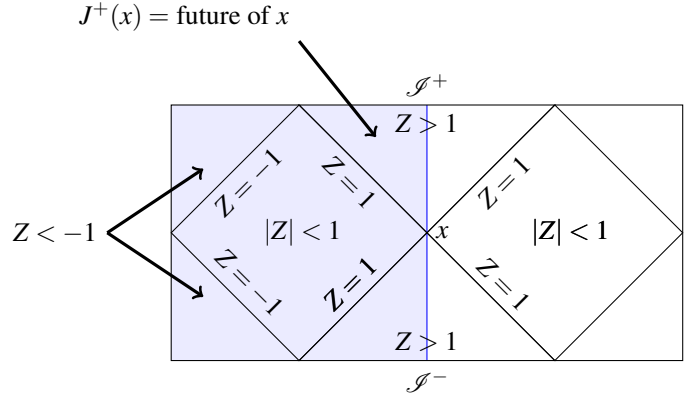

From its definition as a hyperboloid in 5-dimensional Minkowski space,
it is clear that deSitter has the 10-dimensional group $O(4,1)$ as its isometry
group.
Let us now consider Klein-Gordon quantum field on deSitter
spacetime. Since $dS_4$ is globally hyperbolic, we can define the algebra of field observables $\eA(dS_4,g)$
by the general procedure above. For $m^2 > 0$, a globally $O(4,1)$-invariant state exists
called the Bunch-Davies (aka Hartle-Hawking, aka Euclidean) vacuum. To describe this
it is convenient to introduce a function $Z: dS_4 \times dS_4 \to \mr$
by
\ben\label{Zdef}
Z(x,y) = H^{2}\, Y(x) \cdot Y(y)
\een
in terms of the embedding $Y:dS_4 \to \mr^5$ of deSitter space into
five-dimensional Minkowski space. This function is symmetric, deSitter
invariant, and is related to the signed geodesic distance $\sigma$ by the formula
\ben\label{zsigma}
\cos(H \sqrt{\sigma}) = Z \, ,
\een
where the square root is taken to be imaginary for time-like separated points.
The causal relationships between points can be put in correspondence with values of
$Z$; see the conformal diagram, fig.~\ref{dS1}.

In terms of $Z$, the 2-point function of the Bunch-Davies state is~\cite{allen,bros1,bros2}
(in $d$ dimensions)
\ben\label{vac}
W_2(x,y) = \frac{H^{d-2}}{(4\pi)^{d/2}} \frac{\Gamma(-c)\Gamma(c+d-1)}{\Gamma(d/2)} \ {}_1 F_2 \ \left(-c, d-1+c; d/2; \frac{1+Z-it0^+}{2} \right) \, ,
\een
where the dimensionless constant $c$ is defined by
\ben
c = -\frac{d-1}{2} + \sqrt{\frac{(d-1)^2}{4} - \frac{m^2}{H^2}}
\een
and the usual boundary value prescription
has to be applied, with $t=Y^0(x)-Y^0(y)$. It is relatively easy to check
that this 2-point function satisfies the KG-equation in each argument, and it can be verified that its antisymmetric part satisfies \eqref{commu}. It is far less obvious that the 2-point function is positive,
but this be shown using the following rather elegant and non-trivial representation due to~\cite{bros1,bros2}
(assuming for simplicity a ``principal series scalar field''  characterized by $\mu^2 := m^2 - \frac{(d-1)^2}{4} H^2 \ge 0$)
\ben \label{posrep}
W_2(x,y) = {\rm const.}\sum_{l=\pm} \int_{\RR^{d-1}}
\left( Y(x) \cdot \xi({\bf k}, l)
\right)^{c}
\left( Y(y) \cdot \xi({\bf k}, l)
\right)^{\bar c} \, \frac{d^{d-1} {\bf k}}{\sqrt{{\bf k}^2 + \mu^2}} \, .
\een
Here, $\xi({\bf k}, l) \in \RR^{d+1}$ are the
$(d+1)$-dimensional vectors defined by
\ben
\xi({\bf k}, \pm) = (\sqrt{k^2 + \mu^2}, \pm {\bf k}, \pm \mu) \, .
\een
More precisely, $W_2$ is defined again as
the boundary value of the analytic function
obtained by adding to the time coordinate
of $y$ a small positive imaginary part. To check that it is Hadamard,
one may use the relationship between the wave-front set and
distributional boundary values; see appendix~\ref{appA}.

The following statements hold true concerning deSitter invariant states~\cite{allen}:

\begin{itemize}
\item When $m^2>0$, then the Bunch-Davies state is the {\em unique} deSitter invariant, pure, Gaussian,
Hadamard state, although a 1-parameter family of states (`$\alpha$-vacua') exists if the Hadamard condition is dropped.

\item When $m^2 \le 0$, no deSitter invariant state exist, although, as emphasized, in those cases
an infinite set of non-deSitter-invariant Hadamard states still exists. In particular, the algebra $\eA$ may {\em always}
be defined for any value of $m^2$, although when $m^2 < 0$ the $n$-point functions of physically reasonable
states will grow exponentially with time.
\end{itemize}

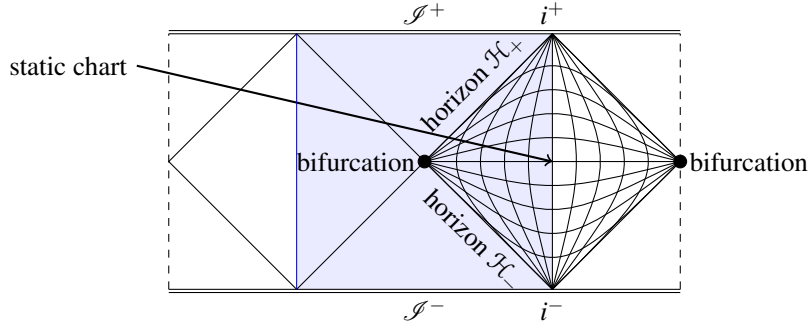
\begin{figure}
\begin{center}
\begin{tikzpicture}[scale=.85, transform shape]
\fill[blue!25!,opacity=.3] (-2,2) rectangle (2,-2);
\draw (-2,2) -- node[above]{${\mathscr I}^+$}  (2,2);
\draw (-4,2.05) --   (4,2.05);
\draw (-2,-2) -- node[below]{${\mathscr I}^-$}  (2,-2);
\draw (-4,-2.05) --   (4,-2.05);
\draw (0,0) node[black,left]{bifurcation};
\draw (0,0) -- (-2,2) -- (-4,0) -- (-2,-2) --   (0,0);
\draw (-2,-2) -- (-4,-2);
\draw[dashed] (-4,-2)  --  (-4,2);
\draw (-4,2) -- (-2,2) ;
\draw (0,0) -- node[black,above, sloped]{horizon ${\mathcal H}_+$}(2,2) -- (4,0) -- (2,-2) -- node[black,below, sloped]{horizon ${\mathcal H}_-$}(0,0);
\draw (2,-2)node[black,below]{$i^-$} -- (4,-2);
\draw[dashed] (4,-2)  -- node[black,right]{bifurcation} (4,2);
\draw (4,0) node[draw,shape=circle,scale=0.5,fill=black]{};
\draw (0,0) node[draw,shape=circle,scale=0.5,fill=black]{};
\draw (4,2)  -- (2,2)node[black,above]{$i^+$};
\draw (0,0) .. controls (2,2) and (2,2) .. (4,0);
\draw (0,0) .. controls (2,1.5) and (2,1.5) .. (4,0);
\draw (0,0) .. controls (2,1) and (2,1) .. (4,0);
\draw (0,0) .. controls (2,0.5) and (2,0.5) .. (4,0);
\draw (0,0) .. controls (2,0) and (2,0) .. (4,0);
\draw (0,0) .. controls (2,-0.5) and (2,-0.5) .. (4,0);
\draw (0,0) .. controls (2,-1) and (2,-1) .. (4,0);
\draw (0,0) .. controls (2,-1.5) and (2,-1.5) .. (4,0);
\draw (0,0) .. controls (2,-2) and (2,-2) .. (4,0);
\draw[->, thick] (-4.5,1.5) node[left] {static chart} -- (2,0);
\draw (2,-2) .. controls (1.5,0) and (1.5,0) .. (2,2);
\draw (2,-2) .. controls (1,0) and (1,0) .. (2,2);
\draw (2,-2) .. controls (.5,0) and (.5,0) .. (2,2);
\draw (2,-2) .. controls (0,0) and (0,0) .. (2,2);
\draw (2,-2) .. controls (2.5,0) and (2.5,0) .. (2,2);
\draw (2,-2) .. controls (3,0) and (3,0) .. (2,2);
\draw (2,-2) .. controls (3.5,0) and (3.5,0) .. (2,2);
\draw (2,-2) .. controls (4,0) and (4,0) .. (2,2);
\draw (2,-2) -- (2,2);
\draw[blue] (-2,2) -- (-2,-2);
\end{tikzpicture}
\caption{
\label{deS2}
Conformal diagram for deSitter spacetime, and the static chart. Again, we are drawing the case $d=2$. 
The case $d>2$ would correspond to the \textcolor{blue}{shaded} square having the bifurcation surface $\cong S^{d-2}$ in the middle.
The vertical boudaries of the shaded square correspond to the north- and south pole of the Cauchy surface $S^{d-1}$. 
}
\end{center}
\end{figure}

In deSitter space, there is a phenomenon reminiscient of the Unruh effect which takes place in the ``static chart;'' see fig.~\ref{deS2}. That chart can be defined as the
intersection of $dS_4$ with a wedge $\{|X_1| > X_0\}$ in the ambient $\RR^5$.
The static chart is again a globally hyperbolic spacetime in its own right, and can also be
defined as the intersection $J^+(i^-) \cap J^-(i^+)$ of two points $i^\pm \in \mathscr{I}^\pm$
which are at the ``same angle''. It can be covered by the coordinate
system $(t, r, \varphi, \theta)$ defined for $t \in \RR, \, 0\le r < H^{-1}$, in which
the line element takes the form
\ben\label{static}
g = - (1-H^2r^2) dt^2 + (1-H^2 r^2)^{-1} dr^2 + r^2 \ (d\theta^2 + \sin^2
\theta d\varphi^2) \ .
\een
It can be seen from this form of the line element that,
within this chart---but of course not
in the full deSitter space---the
metric is static, with timelike Killing field $\xi =
\frac{\partial}{\partial t}$. The corresponding flow
$\varphi_s: t \mapsto t+s$ defines a 1-parameter group of isometries
in the static chart, which correspond to a boost in the $X_0$-$X_1$ plane in the ambient
$\RR^5$. The boundary $\mathcal{H} = \mathcal{H}_+ \cup \mathcal{H}_-$ is formed from two intersecting
cosmological horizons, and is another example of a bifurcate Killing horizon,
with surface gravity $\kappa = H$. The restriction of the Bunch-Davies state to the static chart
is seen to be a {\bf KMS-state} at inverse temperature $\beta = H/2\pi$ by the same
argument as given for the Unruh effect in Rindler spacetime (see \cite{kaywald}). Note that the static orbit corresponding to $r=0$ is a geodesic,
and, by deSitter invariance, any timelike geodesic in deSitter spacetime is an orbit of the static Killing field of some static chart. In this sense, one may say that in the Bunch-Davies state in deSitter spacetime, every freely falling observer will ``feel himself'' immersed in a thermal bath of particles at inverse temperature $\beta = H/2\pi$.

Of particular interest in cosmology is the behavior of the 2-point function of the Bunch-Davies state for large
time-like separation $\tau = \sqrt{-\sigma}$.  Using well-known properties of the hypergeometric function~\eqref{vac},
it is found that the 2-point function behaves as $\e^{-(d-1)H\tau/2}$ for $\tau \gg 1$. This exponential decay reflects the
exponential dispersive effects of fields on deSitter space. It implies that the 2-point function $W^\Psi_2(x,y) = (\Psi, \pi(\phi(x)) \pi(\phi(y)) \Psi)$ of
{\em any} Hadamard state of the form $\Psi := \pi[\phi(f_1) \cdots  \phi(f_n)]\Omega \in \eH, f_i \in C_0^\infty(\M)$ in the GNS-representation 
$(\eH, \pi, \Omega)$ of the Bunch-Davies state 
 approaches that of the Bunch-Davies state [i.e. $W_2(x,y)$, see~\eqref{vac}] when we move $x,y$ towards the distant future keeping the geodesic distance between $x,y$
fixed. Such states are, by construction, dense in $\eH$. The exponential decay corresponds,
physically, to the ``no-hair property'' of deSitter spacetime. As one can show with
considerably more effort, that behavior persists for interacting quantum field theories, see~\cite{h1,h2,marolf1,marolf2}.

\vspace{.5cm}

{\bf c) Hawking effect:}
The algebraic formalism can be used to give a conceptually clear
explanation of the Hawking effect. In fact, there are two closely related, but
distinct, results that are commonly referred to as the ``Hawking effect.''

The first result concerns maximally extended Schwarzschild spacetime (i.e., an ``eternal black hole'').
As is well known, the exterior region, $r>2M$, of
Schwarzschild spacetime
\ben
ds^2 = -(1-2M/r)dt^2 + (1-2M/r)^{-1}dr^2 + r^2(d\theta^2 + \sin^2
\theta d\varphi^2), \quad M > 0,
\een

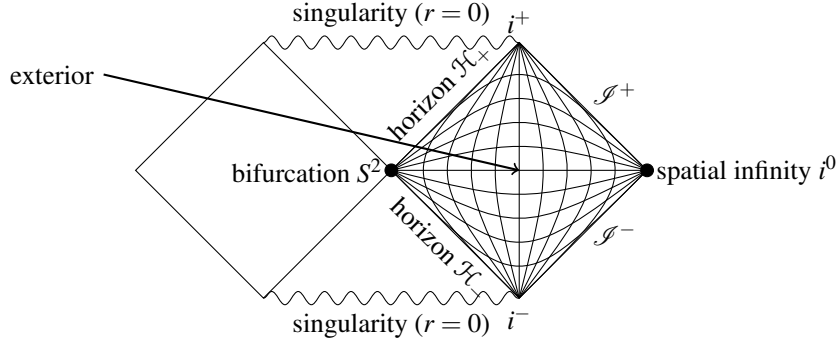
\begin{figure}
\begin{center}
\begin{tikzpicture}[scale=.85, transform shape]
\draw (0,0) node[black,left]{bifurcation $S^2$};
\draw (0,0) -- (-2,2) -- (-4,0) -- (-2,-2) --   (0,0);
\draw (-2,2) decorate[decoration=snake] {-- (2,2)};
\draw (-2,-2)  decorate[decoration=snake] {-- (2,-2)};
\draw (-.0,2.1) node[above]{singularity ($r=0$)};
\draw (-.0,-2.1) node[below]{singularity ($r=0$)};
\draw (0,0) -- node[black,above, sloped]{horizon ${\mathcal H}_+$}(2,2) -- (4,0) -- (2,-2) -- node[black,below, sloped]{horizon ${\mathcal H}_-$}(0,0);
\draw (2,-2 ) node[black,below]{$i^-$};
\draw (4,0) node[black,right]{spatial infinity $i^0$} ;
\draw (3,1.2) node[black,right]{${\mathscr I}^+$} ;
\draw (3,-1) node[black,right]{${\mathscr I}^-$} ;
\draw (4,0) node[draw,shape=circle,scale=0.5,fill=black]{};
\draw (0,0) node[draw,shape=circle,scale=0.5,fill=black]{};
\draw (2,2) node[black,above]{$i^+$};
\draw (0,0) .. controls (2,2) and (2,2) .. (4,0);
\draw (0,0) .. controls (2,1.5) and (2,1.5) .. (4,0);
\draw (0,0) .. controls (2,1) and (2,1) .. (4,0);
\draw (0,0) .. controls (2,0.5) and (2,0.5) .. (4,0);
\draw (0,0) .. controls (2,0) and (2,0) .. (4,0);
\draw (0,0) .. controls (2,-0.5) and (2,-0.5) .. (4,0);
\draw (0,0) .. controls (2,-1) and (2,-1) .. (4,0);
\draw (0,0) .. controls (2,-1.5) and (2,-1.5) .. (4,0);
\draw (0,0) .. controls (2,-2) and (2,-2) .. (4,0);
\draw[->, thick] (-4.5,1.5) node[left] {exterior} -- (2,0);
\draw (2,-2) .. controls (1.5,0) and (1.5,0) .. (2,2);
\draw (2,-2) .. controls (1,0) and (1,0) .. (2,2);
\draw (2,-2) .. controls (.5,0) and (.5,0) .. (2,2);
\draw (2,-2) .. controls (0,0) and (0,0) .. (2,2);
\draw (2,-2) .. controls (2.5,0) and (2.5,0) .. (2,2);
\draw (2,-2) .. controls (3,0) and (3,0) .. (2,2);
\draw (2,-2) .. controls (3.5,0) and (3.5,0) .. (2,2);
\draw (2,-2) .. controls (4,0) and (4,0) .. (2,2);
\draw (2,-2) -- (2,2);
\end{tikzpicture}
\end{center}
\caption{
\label{Sch1}
Conformal diagram of extended Schwarzschild spacetime (`eternal black hole').
}
\end{figure}

may be extended by introducing
the Kruskal coordinates
\ben
U = \e^{-u/4M}, \quad V = \e^{v/4M},
\label{Uu}
\een
where
\ben
u = t - r_*, \quad v = t+r_*,
\een
with $r_* = r + 2M \, \log(r/2M-1)$. In Kruskal coordinates, the line element takes the form
\ben
ds^2 = \frac{32 M^3 \e^{-r/2M}}{r} dUdV + r^2(d\theta^2 + \sin^2
\theta \, d\varphi^2) \, .
\een
By considering arbitrary $U,V$ compatible with $r>0$, one obtains the
maximally extended Schwarzschild spacetime shown in the conformal diagram fig.~\ref{Sch1}. The surfaces $U=0$ and $V=0$ (corresponding to $r=2M$) comprise a bifurcate Killing horizon, ${\mathcal H}^\pm$, of the Killing field $\xi = \partial/\partial t$,
analogous to the bifurcate Killing horizons of the boost Killing field of Minkowski spacetime and the static Killing field of deSitter spacetime. In close analogy with those cases, there exists~\cite{sanders2} a unique~\cite{kaywald}
Hadamard state, $\omega$,
on extended Schwarzschild spacetime
that is stationary i.e., invariant under time-translation automorphisms, $\omega = \omega \circ \alpha_t$. This state is known as the ``Hartle-Hawking vacuum'' and is analogous to the Minkowski vacuum in Minkowski spacetime and to the Bunch-Davies vacuum in deSitter spacetime. By the same argument as in those cases \cite{kaywald}, when restricted to the original Schwarzschild wedge, $r > 2M$, the Hartle-Hawking vacuum is a
KMS state at the Hawking temperature
\ben\label{TH}
T_H = \frac{\kappa}{2\pi} = \frac{1}{8\pi M} \ ,
\een
where $\kappa = 1/4M$ is the surface gravity of the Killing horizon.

Two other states of interest on extended Schwarzschild spacetime are the ``Boulware vacuum''~\cite{boulware}, and the ``Unruh vacuum''~\cite{unruh}.
The Boulware vacuum is defined in the right wedge of extended Schwarzschild spacetime, where it is a ground state with respect to the timelike Killing vector field $\partial_t$.
It is Hadamard in the right wedge, but cannot be extended as a Hadamard state beyond the right wedge, i.e. it would become singular on the past and future horizons. The Unruh
vacuum is defined on the union of the right wedge and the interior of the black hole. In the right wedge, it can be thought of as a KMS state with respect a subalgebra of $\eA$ corresponding
to the modes that are outgoing from the white hole, whereas it the ground state with respect to a subalgebra corresponding to the modes that are incoming from past null infinity.
The Unruh vacuum has been rigorously constructed in~\cite{moretti}, and has been shown to be Hadamard on the union of the right wedge and the black hole regions. It cannot
be extended as Hadamard state beyond the past horizon.

The thermal nature of the Hartle-Hawking state suggests, but does not imply, a second key result: Black holes
formed by gravitational collapse will emit thermal radiation. To analyze this issue, one must consider the
much more physically
relevant case of a spacetime
in which gravitational collapse to a Schwarzschild black hole occurs, rather than the maximally extended
Schwarzschild spacetime considered above. In the case of a black hole formed by collapse,
one can show that if the state of the quantum field is
Hadamard and if it approaches the ground state near spatial infinity, then at late times it contains
quanta of {\em radiation appearing to emanate from the black hole}, distributed according to
a Planck-distribution with temperature~\eqref{TH}.

\begin{figure}
\begin{center}
\begin{tikzpicture}[scale=.8, transform shape]
\shade[top color=gray] (-0.5,4) -- (-0.5,-4.5) .. controls (.4,0) and (.4,0) .. (-.2,4) -- (-.5,4);
\draw (-0.5,-4.5) -- node[above, sloped]{collapsing star} (-0.5,4);
\draw (-0.5,-4.5) -- node[below, sloped]{infinity ${\mathscr I}^-$} (6,2);
\draw (-.2,4) decorate[decoration=snake] {-- (4,4)};
\draw (.2,.2) -- node[black,above, sloped]{horizon ${\mathcal H}_+$}(4,4);
\draw[dashed] (.2,.2) -- (-.5,-.5);
\draw (6,2) -- node[black,above,sloped]{infinity ${\mathscr I}^+$}(4,4);
\draw (-0.5,-4.5) .. controls (.4,0) and (.4,0) .. (-.2,3.9);
\draw (-0.5,4) -- (-.2,4);

\fill[color=blue,opacity=.5] (4,3.5) -- (1.3,.85) -- (2,1.06) --     (4.25,3.25) -- (4,3.5);
\fill[color=blue,opacity=.5] (4,3.5) -- (5.57, 1.93) -- (5.21,1.77) --     (3.75,3.25) -- (4,3.5);
\fill[color=blue,opacity=.2] (4,3) -- (5.21,1.77) --     (2,1.06) -- (4,3);
\fill[color=blue] (4,3.5) -- (4.25,3.25) -- (4,3) -- (3.75,3.25) -- (4,3.5);
%
\draw[blue,thick,directed] (4,3.5) -- (4,4);
\draw[blue,thick,directed] (3,0.3) .. controls (3.9,2) and (3.9,2) ..  (4,3);
\draw[black] (.7,.7) -- (6,2);
\draw[blue, very thick] (1.3,.85) --  node[above,sloped,black]{$$} (5.57,1.89);
\draw (4.5,1.5) node[below]{$\Sigma_0$};
\draw[->, thick] (8,-3) node[right] {support of $f^T_+$} -- (1.7,.8);
\draw[->, thick] (8,-2) node[right] {$f^T$ decaying $\to 0$ as $T \to \infty$} -- (3.25,1.2);
\draw[->, thick] (8,-1) node[right] {support of $f^T_-$} -- (5.4,1.8);
\draw[->, thick] (8,6) node[right] {support of initial data of $F^T$} -- (4.25,3.5);
\draw (-.5,-4.5 ) node[black,below]{$i^-$};
\draw (6,2) node[black,right]{$i^0$} ;
\draw (-.5,-4.5) node[draw,shape=circle,scale=0.5,fill=black]{};
\draw (6,2) node[draw,shape=circle,scale=0.5,fill=black]{};
\draw (4,4) node[black,above]{$i^+$} ;
\draw (4,4) node[draw,shape=circle,scale=0.5,fill=black]{};
\draw (.7,.7) node[black,below right]{${\mathscr S}$} ;
\draw (.7,.7) node[draw,shape=circle,scale=0.5,fill=black]{};
\end{tikzpicture}
\end{center}
\caption{
\label{Sch2}
Conformal diagram of collapsing star spacetime.
}
\end{figure}

We now give some details, following the argument given by
Fredenhagen and Haag~\cite{fred1}.
Let $F^T_{\nu lm}$ be a solution to the Klein-Gordon equation
with smooth initial data of compact support
on the gravitational collapse spacetime
which has $Y_{lm}$ angular dependence and frequency peaked sharply
near $\nu > 0$ (with respect to
the timelike Killing field), and
which corresponds at late times to an outgoing wave reaching null infinity at retarded time
centered about $T$. We normalize $F^T_{\nu lm}$ so that it has unit Klein-Gordon
norm\footnote{The Klein-Gordon inner product between two solutions $F,G$ of compact
support on a Cauchy-surface $\Sigma$ is defined as
$(F,G) = i \int_\Sigma (\bar F \nabla_\mu G - G \nabla_\mu \bar F) n^\mu d S$;
compare~\eqref{KGip}.}. Then, in any state $\omega$, the quantity $\omega(\phi[F^T_{\nu lm}]^* \phi[F^T_{\nu lm}])$
has the interpretation of being the ``number of particles'' in the mode
$F^T_{\nu lm}$ as seen by a distant oberver at late times, as can be seen from
Fock representation formulas (see section 2.2 above) or by considering the
behavior of model particle detectors~\cite{fred4}, \cite{wald2}.
We shall show that if $\omega$ is Hadamard and approaches the ground state near spatial infinity, then
\ben\label{hawking}
\lim_{T \to \infty}
\omega(\phi[F^T_{\nu lm}]^* \phi[F^T_{\nu lm}]) =
\frac{|D_l(\nu)|^2}{\e^{2\pi\nu/\kappa}-1} \, ,
\een
where $D_l(\nu)$ is the
amplitude
for the absorption by the black hole of a mode of angular dependence $Y_{lm}$ and frequency $\nu$.
This is precisely the expected number of particles that one would have for black-body radiation ``emitted'' by the black hole at the Hawking temperature.

To show this, we choose a partial Cauchy surface $\Sigma_0$
intersecting the future horizon $\H^+$ at a $2$-sphere ${\mathscr S} \cong S^2$ outside of the collapsing star (see
fig.~\ref{Sch2}). In the future
domain of dependence $D^+(\Sigma_0)$ of $\Sigma_0$,
the spacetime metric is precisely equal to that of
the Schwarzschild metric. The conditions on the state $\omega$ imply that
\begin{enumerate}
\item[1)] The Hadamard condition in the form~\eqref{localhada} holds for the 2-point function
$W_2(x_1,x_2)$ for $x_1,x_2$ in an open neighborhood of the horizon $\H^+$. Together with the fact that the
coefficient `$u$' in the Hadamard expansion satisfies $u(x_1,x_2) \equiv 1$ for all $x_1,x_2 \in \H^+$, one can infer that
\ben\label{hadamf}
W_2(x_1, x_2)  = \frac{1}{2\pi^2 (\sigma + it0^+)} + \text{lower order terms in $\sigma$ near $\H^+$.}
\een
(Here, lower order refers to the ``scaling degree''; see appendix~\ref{appA}.)

\item[2)] In an open neighborhood of $\Sigma_0$, and for $r_1, r_2 \to \infty$, the
2-point function $W_2(x_1,x_2)$ approaches that of the ground state, i.e., the Boulware vacuum.
\end{enumerate}
The key idea in the derivation
is to replace the solution $F^T_{\nu lm}$ by a test function $f^T_{\nu lm}$
which is supported close to $\Sigma_0$. To do so, we let
$\psi$ be a smooth function which
is equal to $1$ slightly in the future of $\Sigma_0$, and equal to $0$ slightly in the
past of $\Sigma_0$, and we set
\ben
f^T_{\nu lm} = \square_g [\psi F^T_{\nu lm} ] \ .
\een
Then we have $\phi(f^T) = \phi[F^T]$ (see eq.~\eqref{KGip}), and, because at least one derivative must hit $\psi$, $f^T$ is supported near $\Sigma_0$. 
The key property of $f^T$ is that it decays uniformly for large $T$ in any region $r_1 < r < r_2$ where $r_1 > 2M$~\cite{dafermos}.
Thus, in this limit, $f^T$ splits approximately into two parts $f_+^T+f_-^T$ (see fig.~\ref{Sch2}), where the first part, $f_+^T$, is supported close to the horizon $\mathscr S$, and the
second part, $f_-^T$, is supported close to spatial infinity $i^0$ near $\Sigma_0$.
For $f_+^T$, one finds to the appropriate accuracy
for $T \gg 1$:
\ben\label{Fp}
f^T_+(U,V) \sim  D_l(\omega) \ \partial_V
\psi(V)
 \ \partial_U \exp\left( \frac{i\nu}{\kappa} \log(U\e^{\kappa T}) \right)
\een
where for simplicity, we have taken $\psi$ to be a function of $V$ only near $\mathscr S$. Here the logarithmic
dependence on $U$ in the exponent can be traced back to the relationship $U = \exp{\kappa u}$ between ``affine time'' $U$ and ``Killing time''
$u$ on the past horizon ${\mathcal H}^-$ of extended Schwarzschild spacetime (see \eqref{Uu}).
The contribution of $f^T_+$ to
\eqref{hawking} is given by
\ben
\lim_{T \to \infty} \omega( \phi[F_{\nu lm}^T]^* \phi[F^T_{\nu l m}] ) = \lim_{T \to \infty} \int
W_2(x_1, x_2) f^{T}_+(U_1,V_1, {\bf x}_1)^* f^{T}_+(U_2, V_2, {\bf x}_2) \ dv_1 dv_2
\een
On the other hand assumption (2) implies that $f_-^T$ makes no contribution
 in the limit $T \to \infty$, either in its direct terms or its cross-terms
with $f_+^T$.
Since $f_+^T$ is supported increasingly close to ${\mathcal H}^+$, we
may use assumption (1) to approximate $W_2$ with~\eqref{hadamf}. Using furthermore~\eqref{Fp}, 
a short calculation then gives the thermal distribution formula
\eqref{hawking}.

The above argument corresponds closely to Hawking's original derivation~\cite{hawking}, showing that the expected number of particles
seen by a distant observer at late times corresponds to thermal emission by the black hole. In fact, all aspects of
this radiation are thermal~\cite{wald}.  The precise result is that any state that is Hadamard and behaves like a ground near spatial infinity
will approach the Unruh vacuum state, in the sense that the $n$-point functions of the state approach those of
the Unruh vacuum in the exterior at late times.

We have presented the argument in the above manner to emphasize the following points: (i) The derivation of the Hawking effect does
not depend on introducing a notion of ``particles'' near the horizon. (ii) No assumptions need to be made on the initial state other than it is regular (Hadamard) and approaches the ground state near spatial infinity (i.e., there is no ``incoming radiation'' at late times). (iii) The Hawking effect follows from causal propagation of the quantum field outside of the black hole; one does not need to make any assumptions about what is happening inside of the black hole. In particular, any breakdown of known laws of physics in the high curvature regime near the singularity deep inside the black hole should not affect the validity of the derivation. (iv) The details of the collapse are not important; all that matters is that the spacetime metric asymptotically approach the Schwarzschild metric\footnote{The results can be straightforwardly generalized to asymptotic approach to other stationary black hole geometries, with the only significant difference being that, for a rotating black hole, the horizon Killing field will now be a linear combination of a time translation and rotation at infinity; see, e.g. ~\cite{hawking,wald} for a discussion of the Kerr case.} by some sufficiently late ``time'' $\Sigma_0$.

Nevertheless, there is one potentially disturbing aspect of this derivation. For large $T$, it can be seen from \eqref{Fp}
that $f^T_+$ is peaked near $U \sim \exp(-\kappa T)$, and that the locally measured frequency of $f_+^T$---say, as seen by observers who freely fall into the black hole from rest at infinity---diverges as $\exp(\kappa T)$ as $T \to \infty$.  For any reasonable detector frequency $\nu$, this vastly exceeds the Planck frequency for $T \gg 1/\kappa$. We cannot expect QFTCS to be a good approximation to nature on transplackian scales, but the above derivation of the Hawking effect appears to depend upon the the validity of QFTCS at transplanckian frequencies. However, we claim now that this is not actually the case: For a given $F^T_{\nu l m}(r,t,\varphi,\theta)$ at large $T$, instead of doing the analysis on the partial Cauchy surface $\Sigma_0$, we can work on a partial Cauchy surface $\Sigma_1$, which is sufficiently early in time that the approximation leading to \eqref{Fp} still holds, but is sufficiently late in time
that $f^T_+$ on $\Sigma_1$ is not transplanckian. By formulating the detector response at time $T$ as an evolution problem starting from
$\Sigma_1$ rather than $\Sigma_0$, one avoids\footnote{Of course, the Hadamard condition itself concerns arbitrarily short distance singularity structure and thus, in effect, involves transplanckian scales. One might therefore question the validity of QFTCS for arguing that the Hadamard condition is preserved under evolution. However this question could equally well be raised in Minkowski spacetime.} any elements of the derivation that allude to phenomena at transplanckian scales.

The derivation of the emission of thermal radiation by a black hole at the Hawking temperature \eqref{TH}
brought to a culmination a remarkable relationship between certain laws of black hole physics and the ordinary laws of
thermodynamics. It was already known prior to this derivation that classical black holes satisfy mathematical analogs of the zeroth, first, and second laws of thermodynamics~\cite{bardeen},
with mass, $M$, playing the role of energy, $E$; surface gravity, $\kappa$, playing the role of temperature $T$; and horizon area, $A$, playing the role
of entropy, $S$. Even in classical physics, a hint that this relationship might be more than a mathematical analogy is provided by the
fact that $M$ and $E$ are, in fact, the same physical quantity. However, the physical temperature of a classical black hole is absolute zero, thereby spoiling this relationship in classical physics. The fact that, when analyzed from the perspective of QFTCS, black holes have a finite temperature \eqref{TH} proportional to surface gravity strongly suggests that these laws of black hole physics must
actually be the laws of thermodynamics applied to black holes; see \cite{wald2,wald3} for further discussion. In particular, $A/4$ must represent the physical entropy of a
black hole. The ramifications of these ideas continue to be explored.

In the above derivation of thermal emission by a black hole, we considered a fixed, classical spacetime corresponding
to the gravitational collapse of a body to a Schwarzschild black hole. However, the quantum field has a stress-energy observable, $T_{\mu\nu}$ (see the next section), and, in semiclassical gravity, $\omega(T_{\mu\nu})$ should contribute to the right side of Einstein's equation. It is easy to see that $\omega(T_{\mu\nu})$ contributes a positive energy flux to infinity. It follows from conservation of stress-energy together with the approximate stationarity of $\omega$ at late times that $\omega(T_{\mu\nu})$ contributes a corresponding flux of negative energy\footnote{Negative energy fluxes of stress-energy or negative energy densities can occur in quantum field theory even for fields that classically satisfy the dominant energy condition, see e.g. \cite{fewsterqei,ford}.} into the black hole. Consequently, if the black hole is isolated (so that there is no other flux of stress-energy into the black hole), the black hole will slowly lose mass as a result of the quantum field effects. An order of magnitude estimate of the mass loss of the
 black hole can be obtained from the Stefan-Boltzmann law
\ben
\frac{dM}{dt} \sim A T^4 \sim M^2 \frac{1}{M^4} = \frac{1}{M^2}
\een
leading to the prediction that a black hole should ``evaporate'' completely\footnote{Of course, the approximate description leading to this prediction should be valid only when $M \gg M_P$, where $M_P$ denotes the Planck mass ($\sim 10^{-5} \rm{gm}$), but modifications to the evaporation process at this stage (including the possibility of Planck mass remnants) would not significantly alter the discussion below.} in a time of order $M^3$ (in Planck units).

The prediction of black hole evaporation gives rise to an issue that has re-gained considerable attention recently. In the analysis of quantum field theory on the gravitational collapse spacetime shown in fig. 5---where the black hole remains present forever---the field observables in an open neighborhood of any Cauchy surface comprise the entire algebra\footnote{This result, 
sometimes called ``time-slice property'', continues to hold for the enlarged algebra $\eW$ defined in the next section and also for the algebra of interacting fields $\eB_\rI$~\cite{hw3,fred5}.}, $\eA$. Thus, a state that is pure at any initial ``time'' (i.e., a neighborhood of an initial Cauchy surface, say, prior to the collapse) automatically will remain pure at any final ``time'' (i.e., a neighborhood of a final Cauchy surface, say, after the black hole has formed). However, the field observables in a small neighborhood of a {\em partial} Cauchy surface---i.e., a hypersurface, such as $\Sigma_0$ in the above figure~\ref{Sch2}, whose future domain of dependence includes the region exterior to the black hole but not the interior of the black hole---do not comprise all observables in $\eA$ since there will be additional field observables inside the black hole. Furthermore, for any Hadamard state, there always are strong correlations between the field observables at small spacelike separations. In particular, in any Hadamard state, the observable $\phi(f^T)$ on $\Sigma_0$ in our above discussion will be highly entangled with corresponding field observables inside the black hole. Since $\phi[F^T] = \phi(f^T)$ (see \eqref{KGip}), this means that the Hawking radiation flux measured by a distant detector is highly entangled with field observables inside the black hole. If the black hole subsequently completely evaporates as discussed above, the field observables corresponding to the emitted radiation remain entangled with observables inside the (now non-existent) black hole. In particular, the subalgebra of observables that can be measured at late times---after the black hole has evaporated---do not comprise a complete set of observables, and the restriction of the state $\omega$ to this subalgebra is a mixed state. Thus, in the process of black hole formation and evaporation, an initial pure state will evolve to a final mixed state. Such an evolution does not violate any principles of quantum theory or any known laws of physics---indeed, it is
 derived by a straightforward application of QFTCS to a spacetime in which a black hole forms and evaporates---but it is in apparent conflict with ideas suggested by the AdS/CFT correspondence. In any case, there is a widespread belief that evolution from a pure state to a mixed state should not happen, thereby requiring a drastic modification of QFTCS in a low curvature regime, where, a priori, one otherwise would have very little reason to question its validity. In particular, in order to avoid entanglement between observables outside the black hole and inside the black hole, the state must fail to be Hadamard at the event horizon of the black hole, thereby converting the event horizon to a singularity (a ``firewall''). On the other hand, as we have seen above, the Hawking effect itself is crucially dependent upon the state being Hadamard arbitrarily close to (but outside of) the event horizon. Given that the event horizon is not locally determined---i.e., it requires knowledge of the future evolution of the spacetime---it would seem a daunting task for a quantum field to know, with the required precision, exactly when to stop obeying the laws of QFTCS, so that Hawking radiation is maintained but its entanglement with field observables inside the black hole is broken. Nevertheless, there presently is a widespread belief that a quantum field will somehow manage to do this---or that the presently known local laws of physics will be violated near the horizon of a black hole in some other way, so as to maintain the purity of the final state. 

\vspace{.5cm}

{\bf d) Cosmological perturbations:}
We would now like to investigate a massless ($m^2=0$) Klein-Gordon field propagating on
an FLRW-spacetime with flat slices
\ben
g = -dt^2 + a(t)^2 (dx_1^2 + dx_2^2 + dx_3^2) \ .
\een
The isometry group of this spacetime is, for general $a(t)$, the
Euclidean group $E(3)$ acting on the spatial coordinates ${\bf x} \in \RR^3$.
We consider a scale factor of the form
\ben
a(t) =
\begin{cases}
\e^{H_0 t} & \text{for $t \le t_0$,}\\
a_0 \left( \frac{t}{t_0} \right)^{p} & \text{for $t > t_0$,}\\
\end{cases}
\label{inflation}
\een
describing a universe which is expanding exponentially first (inflation), followed
by an era with power-law expansion factor, assumed for simplicity to last
forever. To be precise, we should actually choose scale factor that interpolates smoothly, rather than just continuously,
between the epochs, but for the rough calculation this will not be needed. We could also add
an earlier epoch of power law expansion prior to the exponential expansion, but this will not affect our
results provided that the era of exponential expansion lasts sufficiently long.

We consider a massless quantum Klein-Gordon scalar field in the spacetime \eqref{inflation}. This is a (slight)
simplification of the more physically relevant problem of
starting with a classical solution of the Einstein-scalar-field system with scalar factor
of a form approximating \eqref{inflation}---as would occur if the scalar field ``slowly rolls'' down an
extremely flat potential---and then treating the linearized perturbations of this system
as quantum fields. In that case, the linearized perturbations decouple into ``scalar modes'' and
``tensor modes,'' each of which behave similarly to a scalar field in the background spacetime \eqref{inflation}, see e.g.~\cite{sasaki,mukh} for 
reviews.
Thus, consideration of a scalar field in the background spacetime \eqref{inflation} suffices to derive
the general form of the power spectrum of perturbations resulting from inflation.

Consider a Gaussian, pure, $E(3)$-invariant, Hadamard state of the scalar field on the
spacetime \eqref{inflation}.  The two-point function, $W_2$, of such a state may be described
by a set of mode functions $u_{\bf k} (t, {\bf x}) = \chi_{\bf k}(t) \exp(i{\bf kx})$ as
in eq.~\eqref{modes}.
Let us normalize
the mode functions so that the Wronskian is $i$-times unity, i.e.,
\ben
i = a(t)^{3} \left(
\chi_{\bf k}(t) \frac{d}{dt} \overline\chi_{\bf k}(t) -
\overline \chi_{\bf k}(t) \frac{d}{dt} \chi_{\bf k}(t)
\right) \, .
\een
This condition ensures that eq.~\eqref{wronsk} holds with $X = \RR^3, d\mu(\k) = d^3 \k/(2\pi)^3$
so that eq.~\eqref{modes} indeed defines the 2-point function of a state,
\ben
W_2(t_1, {\bf x_1}, t_2, {\bf x_2}) =
\frac{1}{(2\pi)^3} \int \chi_{\bf k}(t_1) \overline{\chi_{\bf k}(t_2)} \ \e^{i{\bf k}({\bf x}_1-{\bf x}_2)} \ d^3 {\bf k} \ .
\label{inflation2}
\een
A quantity that (partly) characterizes $W_2$
in a Robertson-Walker spacetime is its
``power spectrum,'' $P(t,{\bf k})$, which is defined in terms of the spatial Fourier-transform of the 2-point function at equal time $t$,
\ben
\hat W_2(t, {\bf k}, t, {\bf p}) = (2\pi|{\bf k}|)^3 \delta^3({\bf k}-{\bf p}) \, P(t,{\bf k}) \, .
\een
For an $E(3)$-invariant state, the power spectrum
$P(t,k)$ only depends upon the modulus $k=|{\bf k}|$ of the
wave number (and of course $t$).

We shall now show that for any $t > t_0$ (i.e., after inflation has ended),
the power spectrum $P(t,k)$ for modes for which\footnote{Such modes having wavelength larger than the Hubble radius in the present universe are not of observational interest. However, modes whose wavelength was larger than the Hubble radius at the end of inflation but is smaller than the Hubble radius in the present universe are highly relevant to cosmology.} $k/a(t) \ll 1/R(t)$, where $R$ denotes
the Hubble radius, $R(t) \equiv a(t)/\dot a(t)$, is
approximately given by
\ben\label{ps}
P(t,k) \propto H_0^2 \, .
\een
Thus, the power spectrum is ``scale free,'' with amplitude set by the scale of inflation.
To show this, we observe that
the Hadamard condition fixes the asymptotic behavior of
$\chi_{\bf k}(t)$ for large $|{\bf k}| \to \infty$ (to all(!)
asymptotic orders). In the inflationary epoch ($t < t_0$), the general solution for $\chi_\k$
giving rise to an $E(3)$-invariant state of the massless field is
\ben\label{deSmodes}
\chi_{\bf k}(\eta) =
A_{k} f_k(\eta) +
B_{k} \overline{f}_k(\eta) \ ,
\een
in conformal time $\eta =\int dt/a(t)$,
where
\ben\label{deSm1}
f_k(\eta) = {\rm const.}  \ \eta^{\frac{3}{2}} \, {\rm H}_{-\frac{3}{2}}^{(2)}(k\eta) \ ,
\een
where
${\rm H}_{\alpha}^{(2)}$ denotes a Hankel function,
and the Wronskian condition imposes $|A_k|^2-|B_k|^2=1$. The
Hadamard condition requires $A_k \to 1$, $B_k \to 0$ at large $k$.
If we assume that $A_k \approx 1$, $B_k \approx 0$ for all ``short wavelength
modes'' (i.e., $k/a(t_1) > H_0$) at some time $t_1 < t_0$ during the inflationary
era, then we have $A_k \approx 1$, $B_k \approx 0$ for all modes\footnote{
For $m^2>0$, the analogous modes are obtained by setting the index of
the Hankel function to
$\alpha = (\frac{9}{4}-m^2H^{-2})^{1/2}$. The state with $A_k = 1$, $B_k = 0$
for all $k$ is the deSitter
invariant Bunch-Davies state~\cite{spindel}. However, the choice $A_k = 1$, $B_k = 0$
for $m=0$ would yield an infrared divergence in the two-point function \eqref{inflation2}}  whose
physical wavelength is smaller than $H_0^{-1} \exp[(t_0 - t_1) H_0]$ at the end
of inflation. For $(t_0 - t_1) H_0 \gtrsim 60$, this encompasses all wavelengths
relevant for cosmology.

The mode functions compatible with $E(3)$-invariance
in the power law epoch with $a(t) \propto t^p$ have the form
\ben
\tilde \chi_\k(\eta) = \tilde A_k \tilde f_k(\eta) + \tilde B_k \overline{\tilde f}_k(\eta) ,
\een
where
\ben
\tilde f_k(\eta) =
 {\rm const.}  \ \eta^{\frac{1-3p}{2(1-p)}} \, {\rm H}^{(2)}_{-\frac{1-3p}{2(1-p)}}(k\eta) \, ,
\een
The coefficients $\tilde A_k, \tilde B_k$ are subject to the Wronskian condition
$|\tilde A_k|^2-|\tilde B_k|^2=1$ and are determined by matching the modes
$\chi_\k$ in eq.~\eqref{deSmodes}  to
$\tilde \chi_\k$ at time $\eta_0$. To do this, we first note that, by assumption,
we are considering modes that satisfy $R(t)k/a(t) \ll 1$.
During the power law epoch, we have $a(t) \propto t^p$ with
$p<1$, and $R(t) \propto t$, so the quantity $R(t)k/a(t)$ becomes
even smaller as we go back in time from $t$ to $t_0$. In conformal coordinates, 
$1 \gg R(t)k/a(t) \propto k\eta$, so this
means that the modes $\tilde \chi_\k(\eta)$ of interest are essentially constant (``frozen'')
during the power law epoch for all times before $t$. Thus, we may assume
that $\tilde \chi_\k(\eta)\sim \tilde \chi_\k(\eta_0) = \chi_\k(\eta_0)$ during that epoch, so the power
spectrum at $t$ is essentially the same as that of the state
in deSitter spacetime at time $t_0$ for $k\eta \ll 1$, i.e.,
that obtained using the modes $\chi_\k(\eta_0)$. These modes may then be
approximated by $\chi_\k(\eta) \propto k^{-3/2} H_0$ for $k\eta \ll 1$, thus giving rise to
the desired power spectrum~\eqref{ps}. At this level of approximation, the power spectrum is independent of 
the power $p$ and of the precise nature of the transition period, although the finer properties of
the power spectrum would depend on such details.

For $H_0 \sim 10^{16} \, {\rm GeV}$, the
amplitude of the power spectrum 
is macroscopically large, and it provides an explanation of the observed temperature
fluctuations in the cosmic microwave background as well as of ``structure formation'' in the universe,
i.e., it produces density perturbations appropriate to act as ``seeds'' for the formation of clusters of 
galaxies and galaxies. The fact that, in the
presence of exponential expansion, the short distance quantum fluctuations of fields in the very early
universe can produce macroscopically observable effects in the present universe is one of the most
remarkable predictions of QFTCS.

\section{Perturbative interacting quantum fields}
\label{sec3}

In this section, we extend the algebra $\eA(\M,g)$ of the previous section to an an enlarged algebra
$\eW(\M,g)$ (subsection~\ref{nonlinear}). We then define the algebra of polynomial field observable $\eB_0$ for the free field and we define
time ordered products. These results are then used in subsection~\ref{sec:pertint} to give a perturbative construction of the algebra
of interacting field observables, $\eB_\rI$. In subsection~\ref{sec4}, we give a brief discussion of the additional ideas that are needed
to give similar constructions for gauge fields.

\subsection{Construction of Nonlinear Observables for a Free Quantum Scalar Field}\label{nonlinear}

The construction of the theory of a free quantum field in curved spacetime given in the previous section provides
a mathematically consistent and satisfactory formulation of QFTCS for fields obeying linear equations of motion. However,
even in this case, the theory is incomplete: The observables represented in the algebra $\eA(\M,g)$ consist only of the
smeared fields $\phi(f)$ and their correlation function observables $\phi(f_1) \dots \phi(f_n)$. However, $\eA$ does not include any
observables corresponding to nonlinear functions of the field $\phi$, such as $\phi^2$ or the stress-energy tensor, $T_{\mu\nu}$, of
$\phi$. For this reason alone, one would like to enlarge the algebra of observables $\eA(\M,g)$ to an algebra
$\eW(\M,g)$ that, at the very least, includes smeared versions of all polynomial expressions in $\phi$ and its spacetime derivatives, as well as correlation function observables of these expressions.

There is an additional reason of at least equal importance for wanting to significantly enlarge $\eA$: We wish to formulate
QFTCS for interacting quantum fields. We will discuss a possible approach toward directly doing so in section 5, but,
as an important first step, one would like to give a perturbative construction of an interacting quantum field theory with
Lagrangian of the form
\ben\label{intlag}
\eL = \eL_0 + \eL_1 \ ,
\een
where the Lagrangian,
\ben
\eL_0 = \half ( (\nabla \phi)^2 + m^2 \phi^2 ) \
\een
corresponds to the free Klein-Gordon field, but the interaction Lagrangian, $\eL_1$, contributes nonlinear polynomial terms to the classical equations of motion. A frequently considered example is
\ben
\eL_1 = \lambda \phi^4 \, .
\label{phi4}
\een
In order to define perturbative expressions for quantities of interest in this nonlinear theory, one must define polynomial quantities like ``$\phi^4$''
as (distributional) elements of the algebra of observables of the free theory. One must also define notions of ``time ordered products'' of
$\phi^4$ and other polynomial observables as elements of the free field algebra, since such quantities appear in the formal expressions arising in perturbative expansions. One therefore would like to to enlarge the algebra of observables $\eA(\M,g)$ to an algebra
$\eW(\M,g)$ that includes all time ordered products of all polynomial expressions in $\phi$ and its spacetime derivatives.

The difficulties in defining nonlinear functions of $\phi$ as observables can be understood as arising directly from the distributional nature of $\phi$. One would like to define $\phi^2(x)$ as the pointwise product of $\phi(x)$ with itself, but pointwise products of distributions are, in general, intrinsically ill defined. This is not merely some fussy mathematical point. It is clear from the following simple observation that some sort of nontrivial ``regularization'' will be needed to define nonlinear observables, such as $\phi^2$. For a Hadamard state $\omega$, the two-point function $W_2(x,y) = \omega(\phi(x) \phi(y))$ is a smooth function when
$x$ and $y$ are spacelike separated, so we might expect $\omega(\phi^2(x)) = \lim_{y \to x} W_2 (x,y)$. However, it is easily
seen from \eqref{hadacon} that this limit diverges. In essence, the fluctuations of quantum fields become arbitrarily large at
short distances, making a straightforward construction of nonlinear functions of a quantum field impossible. Nevertheless, a free quantum field should admit nonlinear field observables such as $\phi^2$, and we need to have such observables to define nonlinear QFTCS (perturbatively or otherwise). How are such observables defined/constructed?

It is clear that to define nonlinear observables and time ordered products, we will need a precise characterization of the nature of the singularities of the distributional quantities in $\eA$. The key breakthrough in this regard was made in~\cite{rad1,rad2}---where the equivalence between~\eqref{hadacon} and~\eqref{localhada} was proven---thereby enabling the methods of micolocal analysis to be applied to the renormalization problem, as first done in~\cite{bf1,bf2}.

We ``constructed'' the observable $\phi(f)$ by writing down the free algebra of symbols of this form and factoring by the
relations (1)-(4) of section 2.1, thereby obtaining the algebra $\eA$. This worked because (1)-(4) comprise the complete set of
relations that characterize the linear field observables. We could attempt to define the enlarged algebra $\eW$ by adding symbols
of the form $\phi^2(f)$, etc., to the free algebra and factoring by all of the relations that are satisfied by this enlarged set of observables. However, to do this, we would need to know all of the relations satisfied by the enlarged set of observables, and it is far from obvious, a priori, what this complete list of these relations should be. Remarkably, one can bypass this difficulty and obtain the desired enlarged algebra,
$\eW$, by proceeding as follows~\cite{fred5,hw1}.

The first step is to make a trivial enlargement of $\eA$ to an algebra, $\eA'$, by allowing, in addition to the original elements of the form $\phi(f_1) \dots \phi(f_n)$, elements
corresponding to smearing $\phi(x_1) \dots \phi(x_n)$ with an arbitrary smooth, compact support test function
$F_n (x_1, \dots, x_n)$. In other words, $\eA'$ is obtained by starting the free algebra generated by the formal expressions
\ben\label{fn}
\hat{F}_n = \int \phi(x_1) \dots \phi(x_n) F_n (x_1, \dots, x_n)
\een
and then factoring by the analogs of relations (1)-(4). The replacement of $\eA$ by $\eA'$
makes no essential change to the free field theory.

The second step is to choose a Hadamard state, $\omega$, on $(\M, g)$ and define the normal-ordered product of fields relative to $\omega$ by
\ben
:\phi(x_1) \dots \phi(x_n):_{\omega} = \sum_P \prod_{k \notin P} \phi(x_k) \prod_{(i,j) \in P} [-W_2(x_i, x_j)]
\label{norord}
\een
where $P$ is a collection of disjoint ordered pairs $(i,j)$ such that $i<j$, and where $k \notin P$ denotes the indices $k$ not in
that collection  of pairs.  $W_2$ is as usual the two-point function of the state $\omega$, and both sides are understood to be smeared with a test function $F_n (x_1, \dots, x_n)$.  We will
denote the element of $\eA'$ defined by the smearing of \eqref{norord} with $F_n$ as $:\hat{F}_n:_{\omega}$. For example, we have
\ben
:\hat{F}_2:_{\omega} = \hat{F}_2 - \left( \int_{\M \times \M} F_2 \cdot W_2 \right) \myid  \ .
\een
Using the analog of the commutation relation (condition (3)) on $\eA'$, it may then be seen that elements of the form
$:\hat{F}_n:_{\omega}$ comprise a basis of $\eA'$, i.e., every element of $\eA'$ of the form $\hat{G}_k$ can be expressed as a linear combination of terms of the form $:\hat{F}_n:_{\omega}$ with $n \leq k$. Furthermore, the analog of the commutation relation (condition (3)) on $\eA'$ is effectively encoded by the following rule for calculating products of normal ordered products
\ben
:\widehat{F}_n:_{\omega}  \cdot :\widehat{G}_m:_{\omega} = \sum_{k \le {\rm min}(n,m)}  : \widehat{F_n \otimes_k G_m} :_\omega
\label{wick}
\een
where the $k$ times contracted tensor product is defined by
\bena\label{wick1}
&&F_n \otimes_k G_m(x_1, ..., x_{n+m-2k}) = (-1)^k \frac{n!m!}{k!}
\sum_{ \pi} \int_{\M^{2k}} W_2(y_1, y_2) \ \cdots \ W_2(y_{2k-1}, y_{2k}) \times  \non \\
&& F_n(y_1, y_3, ..., x_{\pi(1)}, ..., x_{\pi(n-k)}) G_m(y_2, y_4, ..., x_{\pi(n-k+1)}, ..., x_{\pi(n+m-2k)}) . 
\eena
The sum $\sum_\pi$ denotes the sum over all permutations $\pi$ of $\{1, ..., n+m-2k\}$ with the
property  $\pi(1)<...<\pi(n-k)$ and $\pi(n-k+1)<...<\pi(n+m-2k)$.
The product formula is a version of ``Wick's theorem.''

The construction of the previous paragraph produces no change whatsoever to the algebra $\eA'$, i.e., it merely rewrites
$\eA'$ in a new basis, with a product law that incorporates the relations already present in $\eA'$.
However, it puts us in position take the major step of enlarging
$\eA'$ to the desired algebra $\eW$ as follows: As noted above, the pointwise product of distributions is, in general, ill defined. However,
there is one significant exception to this statement: As discussed in the Appendix, if $u,v$ are distributions and if their wavefront sets,
$\WF(u)$ and $\WF(v)$, are such that $\WF(u) + \WF(v)$ does not contain a zero cotangent vector, then the pointwise product
$uv$ is well defined in a natural manner. The wavefront set properties of the two-point function $W_{2}$ of a Hadamard state
are such that the products appearing on the right side of \eqref{wick} make sense when the test function $F_n$  is replaced by any
distribution whose wavefront set obeys the restriction
\ben
\WF(F_n) \cap (V_+^n \times V_-^n) = \emptyset \ .
\label{wfw}
\een
Similarly $G_m$ may be replaced by such a distribution.
Thus, we may define a new algebra $\eW$ generated by expressions of the form $:\hat{F}_n:_{\omega}$ with product
law \eqref{wick1}, where $F_n$ is now an arbitrary {\em distribution} of compact support whose wavefront set satisfies
\eqref{wfw}. Since, in particular, for any test function $f$, the distribution $f(x_1) \delta(x_1, \dots, x_n)$ is of the required form, it can be seen that the algebra
$\eW$ includes elements that can be interpreted as representing $:\phi^n(f):_{\omega}$. As we shall see below, the algebra $\eW$ is ``large enough'' to include all polynomial expressions in $\phi$ and its spacetime derivatives as well as all time
ordered products of such expressions. The product law \eqref{wick} provides us with all of the ``relations'' that hold between
these new observables.

The above construction of $\eW$ can be understood as taking a certain ``closure'' of $\eA'$, because distributions $F_n$ satisfying the
wave front set condition~\eqref{wfw} can be approximated, in a suitable topology, by smooth $F_n$'s, which in turn correspond
to elements $: \hat F :_{\omega}$ of $\eA'$. The topology on the $F_n$'s thereby naturally induces a topology on $\eW$, and different
Hadamard states $\omega$ are easily shown to lead to the same topology. It then follows that: (i) The definition of $\eW$ is
independent of the choice of $\omega$. (ii) $\eA'$ is dense in $\eW$. (iii) Any continuous homomorphism on $\eA'$ extends
uniquely to a homomorphism of $\eW$ (since the *-operation and the product,
eq.~\eqref{wick}, can be seen to be continuous in the topology). (iv) Any Hadamard state on $\eA'$ can be extended to a (continuous) state
on $\eW$. Conversely, it can be shown that the restriction to $\eA'$ of any continuous state on $\eW$ yields a Hadamard state on $\eA'$~\cite{sanders2,hollandsruan}. (v) For any Hadamard state, $\omega$, any one-point distribution, $\omega(:\phi^n(\cdot):_\omega)$, is smooth.

The above construction tells us, for any given globally hyperbolic spacetime $(\M, g)$, how to construct the desired
extended algebra of observables $\eW = \eW(\M,g)$. However, it does {\em not} tell us which element of $\eW(\M,g)$ to associate with a given field observable, such as $\phi^2$. In particular, the observable $:\phi^2(f):_{\omega}$ [i.e.,
in our above notation, the observable $:\hat{F}_2:_{\omega}$ with $F_2(x_1,x_2) = f(x_1) \delta(x_1,x_2)$] is {\em not} a reasonable
candidate to represent $\phi^2$ since it depends upon an arbitrary choice of Hadamard state $\omega$. How does one
determine which element of $\eW(\M,g)$ to associate with a given field observable, such as $\phi^2(f)$?

As we shall see, there will be some ``local curvature ambiguities'' in the definition of field observables. However, the
fixing of these ambiguities in the choice of prescription for, say, $\phi^2$ will affect the definition of, say, $\phi^4$. Therefore, in order to properly discuss the ambiguities in the definition of field observables,
it is important that we determine all field observables of interest ``at once.'' The task at hand can then be formulated
as follows. Consider the space of {\it classical} polynomial expressions,
$\O[(\M, g);\phi]$, in the scalar field and its derivatives defined on arbitrary globally hyperbolic spacetimes $(\M,g)$,
i.e., for each globally hyperbolic spacetime $(\M,g)$, $\O(\phi)$ is a polynomial expression in $\phi$ and its derivatives
on $(\M,g)$ (with coefficients that may depend upon spacetime point). We are interested in the $\O$ that are {\it local and covariant}
in the sense that if $\psi:\M \to \M'$ is an isometric embedding (i.e., $\psi^* g'=g$) that also preserves the causal structure---so that if $x_1$ and $x_2$ cannot be connected by a causal curve in $\M$, then $\psi(x_1)$ and $\psi(x_2)$ cannot
be connected by a causal curve in $\M'$---then $\O[(\M, g),\phi]$ must satisfy
\ben
\psi^* \O[(\M',g'), \phi] = \O[(\M,g),\psi^* \phi] \, .
\een
Thus, for example, $\phi^2$ and $R^{\mu\nu} \nabla_\mu \phi \nabla_\nu \phi$ (where $R_{\mu\nu}$ is the Ricci curvature) are local and covariant expressions,
whereas $\phi \int_\M R$ and $v^\mu \nabla_\mu \phi$ (where $v^\mu[\M]$ is a vector field independent of $g$ introduced on each manifold $\M$) are not local and covariant expressions. Note that one cannot tell if an expression $\O$
is local and covariant unless it is defined for {\it all} globally hyperbolic spacetimes.
Let $\P(\M)$ denote the space of classical local and covariant polynomial expressions in $\phi$.
By the Thomas replacement theorem~\cite{iyerwald}, any $\O \in \P(\M)$ may only depend on derivatives of the metric in the form of the Riemann tensor and its (symmetrised) covariant derivatives $\nabla_{(\mu_1} ... \nabla_{\mu_k)} R_{\alpha\beta\gamma\delta}$. Furthermore, at any
point $x \in \M$, $\O$ may only depend on fields $\phi$ and their (symmetrised) covariant derivatives $\nabla_{(\mu_1} ... \nabla_{\mu_k)} \phi$ evaluated {\em at the point} $x$. This result allows us to assign a dimension to terms in $\P (\M)$, by assigning $\nabla_\mu$ to have dimension $1$, $\phi$ to have dimension 1, and $R_{\mu\nu\sigma\rho}$ to have dimension 2.

The question of how to determine which element of $\eW(\M,g)$ to associate with a given field observable, $\O$,
can now be reformulated as how to define a suitable map
$\eQ$ (``for quantization'') from $\P(\M)$ to distributions valued in $\eW$. Our strategy for obtaining $\eQ$ is to write down the conditions that we want $\eQ$
to satisfy, and then determine existence and uniqueness. One obvious condition on $\eQ$ is that it map the classical field expression $\phi \in \P$ to the $\eW$-valued distribution $\phi$. A key condition that $\eQ$ should satisfy
on general elements of $\P$ is that the definition of the quantum field observables should be ``local and covariant.'' This implements
the basic requirement, motivated by general relativity, that there be no ``background structure''
appearing in the formulation of QFTCS
other than the spacetime manifold, $\M$, and the metric, $g$, and that the laws of
QFTCS be local in the metric and the quantum fields. 

To define the local and covariant requirement on $\eQ$,
we first note that our construction of $\eW(\M,g)$ is local and covariant in the following sense~\cite{bfv}:
Let $(\M,g)$ and $(\M',g')$
be globally hyperbolic spacetimes and suppose
that there is a causality preserving isometric embedding
$\psi:\M \to \M'$. Then there exists a corresponding canonical homomorphism $\alpha(\psi): \eA'(\M,g) \to \eA'(\M',g')$
defined by mapping $\hat{F}_n \in \eA'(\M,g)$ as in eq.~\eqref{fn} to the
element of $\eA'(\M',g')$ obtained by replacing the test function $F_n$ by $\psi_* F_n$, (viewing $\phi$
now as the algebra-valued distribution on $\M'$). Since $\eA'$ is dense in $\eW$, $\alpha(\psi)$ uniquely extends to a homomorphism
from $\eW(\M,g)$ to $\eW(\M',g')$, which we also denote as $\alpha(\psi)$. 

The desired local and covariant
condition on the
map $\eQ$ is that for all test functions $f$ on $\eM$ and all causality preserving isometric embeddings
$\psi: \eM \to \eM'$, we have~\cite{bfv}
\ben
\alpha(\psi) [\eQ(\O)(f)] = \eQ(\O) (f \circ \psi^{-1}) \, .
\label{lc}
\een
Note that one cannot tell if the map $\eQ$ satisfies this local and covariant condition without knowing how $\eQ$ is defined on {\em all} globally hyperbolic spacetimes.

If $\eQ$ satisfies \eqref{lc}, then any $\eW(\M,g)$-valued distribution (defined on all globally hyperbolic $(\M,g)$) lying in the image of $\eQ$, will be called a {\it local and covariant quantum field}. Thus, $\eQ(\phi)$ (which we have been denoting---and will continue to denote---as $\phi$) is local and covariant. On the other hand, it is not difficult to show
that for {\em any} assignment of a Hadamard state $\omega$ to each globally hyperbolic spacetime, the quantum field $:\phi^2:_{\omega}$
is {\em not} local and covariant. In essence, ``preferred states'' cannot be locally and covariantly constructed from the
metric, so $\omega$ provides additional, unwanted background structure. Thus, our condition \eqref{lc} on $\eQ$ precludes
the definition $\eQ(\phi^2) = :\phi^2:_{\omega}$ for any assignment of $\omega$ to $(\M,g)$.

In addition to \eqref{lc}, we can make a list of other properties that $\eQ$ should satisfy, including (1) appropriate commutation
relations of $\eQ(\O)$ with $\phi$, (2) appropriate continuous/analytic behavior under continuous/analytic variation of the metric, and (3) scaling of $\eQ(\O)$ that, up to logarithmic terms, agrees with classical scaling under scaling of the metric. We refer the reader to \cite{hw1} for a complete and precise statement of the required properties.

It was proven in \cite{hw1} that the map $\eQ$ defined by
\ben
\eQ(\O) = :\O:_{H_N}
\een
satisfies all of these properties
where the ``Hadamard normal ordering'' operation $: \ \cdot \ :_{H_N}$ is defined by the same formula as \eqref{norord}, with the two-point function $W_{2} (x_1, x_2)$ of the Hadamard state $\omega$ replaced by a locally and covariantly constructed Hadamard distribution $H_N(x_1, x_2)$ (see~\eqref{localhada}). The expansion order $N$ must be chosen to be greater
than the highest derivative in $\Phi$, but is otherwise arbitrary.
Uniqueness up to addition of suitable ``local curvature terms'' can then be proven recursively, using the commutation relations. Details of the existence and uniqueness proofs can be found in \cite{hw1}. A precise statement of the uniqueness result appears below as a special case ($n=1$) of Theorem \ref{Tuniqueness}.

We define the {\it extended algebra of field observables}, $\eB_0(\M,g)$, for the free field to be the subalgebra of
$\eW(\M,g)$ generated by elements lying in the image of $\eQ$, i.e.,
\ben\label{B0def}
\eB_0 \equiv {\rm alg} \{ \eQ(\O) (f) \mid \O \in \P(\M,g), \ \ f \in C^\infty_0(\M) \} \ .
\een
In particular, $\eB_0$ includes all field observables corresponding to local and covariant polynomial expressions in $\phi$ and its derivatives,
such as the stress-energy tensor, $T_{\mu\nu}$. All Hadamard states, $\omega$, yield states on $\eB_0$
and all one-point distributions in Hadamard states, such as the expected stress-energy tensor $\omega(\eQ(T_{\mu\nu})(\ \cdot \ ))$,
are smooth.

\vspace{.5cm}
{\bf Example: Casimir effect in KG-theory on $\M = \RR \times \TT^3$:}
We now illustrate the definition of composite operators
by providing a calculation of the Casimir effect.
We consider the spacetime $\M = \mr \times \TT^3$ equipped with the
flat metric induced from Minkowski space via the identification $\TT^3 = (\RR/2\pi L)^3$, and
we assume that $m^2 > 0$. The ground state $\omega_0$ for the KG-field can e.g. be
found from the corresponding positive frequency modes $u_{\bf k}(t, {\bf x}) = [(2\pi L)^{3/2} \ 2\omega_{{\bf k}/L}]^{-1} \, \e^{-i\omega_{{\bf k}/L}t + i{\bf kx}/L}$,
where ${\bf k} \in \ZZ^3$. According to~\eqref{modes}, the two-point function is found to be
\ben
\label{casimir2}
\begin{split}
\omega_0(\phi(x_1) \phi(x_2)) =&
\frac{1}{(2\pi L)^3} \sum_{{\bf k} \in \ZZ^3}
\frac{
\e^{-i\omega_{{\bf k}/L}(t_1-t_2)}
}
{
2\omega_{{\bf k}/L}
} \ \e^{i{\bf k}({\bf x}_1-{\bf x}_2)/L} \\
=& \frac{1}{(2\pi)^3} \sum_{{\bf n} \in \ZZ^3}  \int
\frac{
\e^{-i\omega_{{\bf p}}(t_1-t_2)}
}
{
2\omega_{\bf p}
} \
\e^{
i{\bf p}({\bf x}_1-{\bf x}_2 + 2\pi L {\bf n})
} \, d^3 {\bf p} \\
=& \sum_{{\bf n} \in \ZZ^3} \hat D^\infty_+(t_1-t_2 - i0^+,
{\bf x}_1 - {\bf x}_2 + 2\pi L {\bf n})
\end{split}
\een
where $D^\infty_+$ was defined above in eq.~\eqref{Ddef} and a hat denotes a Fourier transform in the spatial variables; explicitly
\ben
\hat D^\infty_+(t,{\bf x}) = \frac{m}{4\pi^2}
\frac{
{\rm K}_1\left( m\sqrt{-t^2 + {\bf x}^2+it0^+}\right)
}
{
\sqrt{-t^2 + {\bf x}^2+it0^+}
} \ ,
\label{vacst}
\een
where ${\rm K}_1$ is a modified Bessel function.
To go to the second line in eq.~\eqref{casimir2}, we used the Poisson summation formula, which is applicable here in the sense of distributions in $t$.
The above 2-point function defines a Gaussian state on $\eB_0$.
We wish to define the composite field $\phi^2$ in this theory and compute its expectation value in the state $\omega_0$.
According to the prescription given above, we need to know the Hadamard parametrix $H_N$ (see~\eqref{localhada}) on the spacetime $\M$. Since
$\M$ is locally the same as Minkowski space, it is actually locally identical to that in Minkowski space. In fact, we have,
with $x^2 = -t^2 +{\bf x}^2, e=(1,0,0,0)$, and $N=0$:
\ben
H_0(x_1, x_2) = \frac{1}{2\pi^2} \left( \frac{1}{(x_1-x_2-ie0^+)^2} + \frac{1}{8} m^2 \log (x_1-x_2-ie0^+)^2  \right) \ ,
\een
which is equal to the singular terms in $\hat D^\infty_+(t_1-t_2-it0^+,
{\bf x}_1 - {\bf x}_2)$. The ``local and covariant Wick power'' is then $\phi^2 = :\phi^2:_{H_0}$. Its expectation value is found using the definition
of the normal ordering prescription:
\ben
\omega_0(\phi^2(x)) = \lim_{t \to 0} \left( \omega_0(\phi(x+t\xi) \phi(x-t\xi)) - H_0(t\xi,-t\xi) \right) \ .
\een
Inserting the explicit formula for the 2-point function of $\omega_0$ yields
\ben\label{vev}
\omega_0^{}(\phi^2(x)) = -\frac{1}{16\pi^2} m^2 \log m^2 + \sum_{{\bf n} \in \ZZ^3, {\bf n} \neq {\bf 0}} \hat D^\infty_+(0, 2\pi L {\bf n})  
\een
where it should be noted that the term with ${\bf n}={\bf 0}$ is cancelled, up to $-\frac{1}{16\pi^2} m^2 \log m^2$, by $H_0$, 
as can be seen from the asymptotic expansion of
$\hat D^\infty_+$ in eq.~\eqref{vacst}. As a consequence of the exponential decay of $K_1$, the sum on the right side converges
rapidly for $m^2 > 0$ and can be efficiently evaluated numerically. The sum is practically zero for $2\pi Lm \gg 1$, i.e.
for small Compton wavelength compared to the identification scale. A precursor of this derivation of the Casimir effect in QFTCS was given in~\cite{kaycasimir}.

The definition of $\phi^2$ is not unique---in fact,
according to the previous discussion, the alternative definition $\phi^2 \to \phi^2 + (c_1 R + c_2 m^2) 1$
would be equally acceptable, where $c_1,c_2$ are any real constants, and where $R$ is the scalar curvature.
Since $R=0$ in the present context, the ambiguity consists in adding to $\phi^2$ the operator $c_2
m^2 \ 1$, which changes $\omega_0(\phi^2(x))$ by $c_2 m^2$. We can fix this ambiguity by fixing the expectation
value for any given but fixed $m=m_0$, $L=L_0$. If one is considering a scalar field of mass $m_0$, the usual choice
of $c_2$ would be $c_2 = \log m_0^2/16 \pi^2$, so that\footnote{Note that according to this prescription, the 
expectation value at any other $m \neq m_0$ would then be different from zero for $L \to \infty$.} $\lim_{L \to \infty} \omega_0(\phi^2) = 0$. Once $c_2$ has been
chosen, the expectation value of $\phi^2$ is uniquely determined for any other $m,L$,
and in fact for any other state, such as a finite temperature state. 
A similar calculation can be performed for the energy density operator, and the corresponding 
expectation value is called the `Casimir energy'.

\vspace{.5cm}
The situation with regard to defining time ordered products of polynomial expressions---as needed to define perturbative interacting quantum field theory---is similar to the above problem of defining $\eQ$,
although it is considerably more complicated. The time ordered product of $\O_1(x_1), \dots, \O_n(x_n)$ should be
linear in each $\O_i$, so it is natural to view time order products in $n$ factors as a map, $\T_n$, from the $n$-fold tensor product
$\P^{\otimes n} \equiv \P \otimes ... \otimes \P$ into distributions in $n$ variables valued in $\eW$.
We will denote time ordered products by $\T_n(\O_1(x_1) \otimes \dots \otimes \O_n(x_n))$.
For one factor, we define
\ben
\T_1 (\O) = \eQ(\O) \, .
\een
In order to make the notation less cumbersome, we will
omit writing $\eQ$ (as we have already been doing for the quantum field $\phi$) when it is clear that we are referring to the quantum field $\eQ(\O)$ rather than the classical polynomial $\O$, so, e.g., we will write $\T_1(\O) = \O$.

We would like to define $\T_n$ by ``time ordering'' the product of fields, so, e.g., for
$n=2$ we would like to set
\ben\label{cases}
\T_2(\Phi_1(x_1) \otimes \Phi_2(x_2)) =
\begin{cases}
\Phi_1(x_1) \Phi_2(x_2) & \text{if $x_1 \notin J^-(x_2)$}\\
\Phi_2(x_2) \Phi_1(x_1) & \text{if $x_2 \notin J^-(x_1)$}~.
\end{cases}
\een
(If $x_1$ and $x_2$ cannot be connected by a causal curve, then $\Phi_1(x_1)$ and $\Phi_2(x_2)$ commute, so
either formula may be used.) The problem is that $\Phi_i(x_i)$ are distributions, so \eqref{cases} provides
a definition of $\T_2(\Phi_1(f_1) \otimes \Phi_2(f_2))$ only when the supports of the test functions $f_1$ and $f_2$ satisfy
the relations ${\rm supp} f_1 \cap J^-[{\rm supp} f_2] = \emptyset$ or ${\rm supp} f_2 \cap J^-[{\rm supp} f_1] = \emptyset$. It is not difficult to see that this enables us to straightforwardly define $\T_2(\Phi_1(f_1) \otimes \Phi_2(f_2))$ whenever ${\rm supp} f_1 \cap {\rm supp} f_2 = \emptyset$.
However, \eqref{cases} makes no sense when ${\rm supp} f_1 \cap {\rm supp} f_2 \neq \emptyset$. Thus, we
must extend the distribution \eqref{cases} to the ``diagonal'' $x_1=x_2$. This may seem like a relatively
trivial problem, but the extension of the definition of general time ordered products $\T_n(\Phi_1(x_1) \otimes \dots \otimes \Phi_n(x_n))$
to the ``total diagonal'' $x_1=\dots=x_n$ is the main problem of renormalization theory
in flat and curved spacetime~\cite{eg,bs}. We refer to a definition/construction of $\T_n$ as a ``renormalization scheme.''

We proceed, as in the construction of $\eQ$ above, by writing down a list of properties that $\T_n$
should satisfy. We have already required $\T_1 = \eQ$. We require $\T_n$ to satisfy the appropriate generalization
of \eqref{cases} (involving lower order time-ordered products) away from the total diagonal.
As in the case of $\eQ$, we require $\T_n$ to be locally and covariantly defined, to satisfy
appropriate commutation relations with $\phi$, to have appropriate continuous/analytic dependence on the metric, and to have appropriate scaling behavior (up to logarithmic terms) under scaling of the metric. We also require commutation of $\T_n$ with derivatives. Finally, we impose a number of additional conditions on $\T_n$, specifically, a
microlocal spectrum condition, a ``unitarity'' condition,
and conditions that guarantee that the perturbatively defined interacting field (see subsection~\ref{sec:pertint} below) (i) satisfies the
interacting field equation and (ii) has a conserved stress-energy tensor. The microlocal spectrum condition is given by
eq.~\eqref{msc} of appendix~\ref{appA}.
The unitarity condition is a version of the ``optical theorem'' and it
guarantees that the algebra of interacting fields has representations on a positive definite Hilbert space. It is formulated as follows:
Let $\bar \T_n(\otimes_i \O_i(x_i)) =
[\T_n(\otimes_i \O_i(x_i)^*)]^\dagger$ be
the \lq anti-time-ordered' product\footnote{It can be shown that the anti-time-ordered product in satisfies the causal factorization property with the reversed time-orientation.}.
Then we require
\begin{equation}
\bar \T_n \big( \otimes_{i=1}^n \O_i(x_i) \big) =
\sum_{I_1 \, \sqcup \, ...\, \sqcup \, I_j \, =\, \underline{n}} (-1)^{n + j} \; \T_{| I_1 |} \big( \otimes_{i \in I_1} \O_i(x_i) \big) \; \dots \; \T_{|I_j|}\big(\otimes_{j \in I_j} \O_j(x_j) \big)~,
\label{atop}
\end{equation}
where the sum runs over all partitions of the set $\underline{n} = \{1, \dots, n\}$ into pairwise disjoint subsets $I_1, ..., I_j$.
The condition that guarantees satisfaction of the field equation for the interacting field is a version of the ``Schwinger Dyson equation'', and reads
\ben\label{dyson}
\T_{n+1}\bigg(
(\square_g - m^2) \phi(y) \otimes \bigotimes_{i=1}^n \O_i(x_i)
\bigg)
= i \sum_{k=1}^n
\T_n
\bigg(
\O_1(x_1) \otimes \cdots \frac{\delta \O_k(x_k)}{\delta \phi(y)} \otimes \cdots \O_n(x_n) \bigg) \ .
\een
The condition guaranteeing conservation of the interacting field stress-energy is discussed in \cite{hw4}. We refer the reader
to \cite{hw1}, \cite{hw2}, and \cite{hw4} for a more complete and extensive discussion of all of the conditions imposed on $\T_n$.

It was proven in~\cite{hw2}---key parts of which were based on~\cite{bf2}---that there exists a definition
of $\T_n$, that satisfies all of the above conditions. Furthermore, $\T_n$ is unique up to ``appropriate local and covariant counterterms.'' To explain this freedom in the choice of $\T_n$, we must introduce a considerable amount
of additional notation:  We denote by $\P(\M^n)$ the space of all
distributional local, covariant functionals of $\phi$ (and its covariant
derivatives), of $g$, and of the Riemann tensor (and its covariant derivatives),
which are supported on the total diagonal (i.e. of delta-function type). Let $F= \lambda \int f  \O$ be an integrated local
functional $\O \in \P(\M)$, and formally combine the
time-ordered functionals into a generating functional
written
\ben
\T(\exp_\otimes \left( F \right) ) = \sum_{n=0}^\infty \frac{1}{n!} \T_n(F^{\otimes n}) \in \eW \rf{\lambda}~,
\label{wlambda}
\een
where $\exp_\otimes$ is the
standard map from the vector space  of local actions to the tensor algebra
over the space of local action functionals, and $\eW \rf{\lambda}$ denotes the algebra of formal
power series expressions in $\eW$. Let $\D$ denote a
hierarchy $\D_n$ of linear functionals
\ben\label{Dndef}
\D_n: \P(\M) \otimes \dots \otimes \P(\M) \to
\P(\M^n) \, .
\een
We similarly write $\D(\exp_\otimes (F))$ for the corresponding generating functional
obtained from $\D_n$. We then have the following theorem:

\begin{theorem}\label{Tuniqueness}\cite{hw1} \cite{hw2}, \cite{hw3}, \cite{bf2}
(Uniqueness) If
$\T_n$ and $\hat \T_n$ are two different renormalization schemes, both satisfying our conditions, then they are related by
\ben
\label{unique1}
\hat \T ( \exp_\otimes \left( iF \right) ) = \T \left(
\exp_\otimes \left[ iF + i \D({\rm exp}_\otimes  F  ) \right] \right)   \, .
\een
for any $F = \lambda \int f \O  $, $\O \in \P(\M)$ and $f \in C^\infty_0(\M)$. The functionals $\D_n$ are maps as specified in~\eqref{Dndef} and satisfy:
\begin{enumerate}
\item[(i)] $\D(\e^F_\otimes)=O(\hbar)$ if we reintroduce $\hbar$.
\item[(ii)] Each $\D_n$ is locally and covariantly constructed from $g$.
\item[(iii)] Each $\D_n$ is an analytic functional of $g$.
\item[(iv)]  Each $\D_n(\O_1(x_1) \otimes \dots \otimes \O_n(x_n))$ is a distribution that is supported on the total diagonal ($=$ \lq contact term' $=$ \lq delta-function type').
\item[(v)] The maps $\D_n$ are real.
\item[(vi)] Each $\D_n$ is symmetric.
\item[(vii)] Each $\D_n$ satisfies the natural dimension constraint.
\item[(viii)] Derivatives can be pulled into $\D_n$. This restricts the ambiguities of time-ordered products of fields which are total derivatives.
\end{enumerate}
Conversely, if $\D_n$ has these properties, then any $\hat \T$ given by~\eqref{unique1} defines a new renormalization scheme satisfying our conditions.
\end{theorem}

The expressions $\D_n$ corresponds to
the ``counterterms'' that characterise the difference between the
two renormalization schemes. For the case $\O =  \phi^4$, relevant for the interaction Lagrangian \eqref{phi4}, the
expression
\ben
\delta \eL(y) = \sum_{n \ge 0} \frac{\lambda^n}{n!} \int_{\M^{n-1}} \D_n(\phi^4(y) \otimes \phi^4(x_1) \otimes \cdots \phi^4(x_{n-1})) dv_1 ... dv_{n-1}
\een
corresponds to the finite counterterms in the Lagrangian that arise from a change of the renormalization scheme, with the
$\lambda^n$-term corresponding to the contribution at $n$-th order in perturbation theory. From the properties of
the maps $\D_n$, these will correspond to linear combinations of all possible local covariant expressions of dimension 4, i.e.
the monomials already present in the original Lagrangian $\eL$, together with linear combinations of $R\phi^2$ and  the `$\CC$-number'
terms $m^4, m^2 R, R^{\mu\nu}R_{\mu\nu}, ...$. For other $\O$'s,
the $\D_n$ likewise characterize the ambiguities of composite interacting fields, which are defined below.
The following examples illustrate the allowed ambiguities in the choice of renormalization scheme.

\medskip
\noindent
{\bf Example:}
For $n=1$,
and $F=\int f \phi^2 \dv$, the formula~\eqref{unique1} gives
\ben
\hat \T_1(\phi^2(x)) = \T_1(\phi^2(x)) + \D_1(\phi^2(x)) \ .
\een
and the ambiguity  must have the form
\ben
\D_1(\phi^2(x)) = b_1 m^2 + b_2 R(x) \ ,
\een
where $b_1,b_2$ are real numerical constants.
At the next order $n=2$, the formula~\eqref{unique1} gives (assuming for simplicity that $\D_1 = 0$)
\ben
\hat \T_2(\phi^2(x) \otimes \phi^2(y)) = \T_2(\phi^2(x) \otimes \phi^2(y)) +
\T_1(\D_2(\phi^2(x) \otimes \phi^2(y)))~.
\een
Since the
scaling degree (see appendix~\ref{appA}) of the delta function is $4$ and the dimension of $\phi$ is 1, 
the conditions on $D_2$ stated in Theorem \ref{Tuniqueness} imply that it must take the form
\ben
\D_2(\phi^2(x) \otimes \phi^2(y)) = c_0 \, \delta(x,y)~,
\een
for some real constant $c_0$. Similarly
\ben
\D_2(\phi^3(x) \otimes \phi^3(y)) = c_1 \delta(x,y) \phi^2(y) + (c_2 R + c_3 \square_g + c_4 m^2) \delta(x,y) \ ,
\een
because the scaling degree of $\square_g \delta(x,y)$ is 6, and the dimension of $R$ is 2. Condition 
(viii) of the theorem implies that $c_1 = 9c_0$.
An example with $n=3$ factors is
\ben
\D_3(\phi^2(x) \otimes \phi^3(y) \otimes \phi^3(z)) = c_5 \delta(x,y,z) \ ,
\een
because the scaling degree of the delta function with three arguments is 8.

\subsection{The algebra $\eB_\rI$ of interacting fields}\label{sec:pertint}

We are now in a position to perturbatively define the composite fields for the interacting field theory
described by the Lagrangian density (\ref{intlag}), and the algebra
$\eB_\rI$ of which they are elements. Recall that we have already defined 
$\eB_0$, which corresponds to the case when interaction is turned off,
$\lambda = 0$; see eq.~\eqref{B0def}. 
In the following, we denote the composite fields of the free theory by 
$\O_0(f) \equiv \eQ(\O)(f)$ (i.e., we add the subscript ``$0$"), in order to distinguish them from the corresponding composite fields of the interacting theory, $\O_\rI(f)$,
which we will define below. 

The basic idea to construct the interacting fields, $\O_\rI(f)$, is to initially ``turn off'' the interaction at some finite time in the past, so that $\O_\rI = \O_0$ at sufficiently early times. We then evolve $\O_\rI$ forward in time into the region where the interaction is fully turned on. Finally, we take a limit where the ``turn-on time'' of the interaction is arbitrarily far in the past. However, this last step is problematical because this limit need not exist. This difficulty will be overcome by modifying the limit so that rather than fixing $\O_\rI = \O_0$ in the asymptotic past, we fix $\O_\rI$ in regions of increasing size in the interior of the spacetime.

To implement this strategy to define $\O_\rI(f)$, we choose a cutoff function, $\theta$, of compact
support on $\M$ which is equal to 1 on an open neighborhood of some globally hyperbolic open region $V$ with
the property that $\Sigma \cap V$ is a Cauchy surface for $V$ for some
Cauchy surface $\Sigma$ in $\M$. For $F = \lambda \int_\M \O(x) f(x) \ dv_g$ with $f \in C^\infty_0(\M)$,
we define the {\bf local $S$-matrix} to be
\ben
S(F) = \T(\exp_\otimes (iF)) \equiv \sum_{n \ge 0} \frac{i^n}{n!} \T_n(F \otimes \cdots \otimes F) \ .
\label{locs}
\een
We define the {\bf relative $S$-matrix}
with respect to the cut-off interaction, $\eL_1(\theta) = \int_\M \theta \eL_1$, by
\begin{equation}
\label{rels}
S_{\eL_1(\theta)}(F) = S(\eL_1(\theta))^{-1} S(F+\eL_1(\theta)) \ .
\end{equation}
Then the interacting field, for the interacting theory
with cutoff interaction $\eL_1(\theta)$ corresponding to $\O$ is defined by \cite{bs}
\begin{eqnarray}
\O(f)_{\eL_1(\theta)} &\equiv&
\frac{1}{i} \frac{d}{dt} S_{\eL_1(\theta)}( t\O(f) )
\bigg|_{t=0}.
\label{intfield}
\end{eqnarray}
Equations~\eqref{locs}-\eqref{intfield} are to be understood as formal series
expressions that define the interacting field to any finite order in perturbation theory;
no convergence properties are claimed. Note that
the definition of $\O(x)_{\eL_1(\theta)}$ has been adjusted
so that it coincides with the corresponding free
field $\O(x)$ before the interaction is ``switched on,''
can be seen explicitly by expressing it
in terms of ``totally retarded products''~\cite{hw3,bf2,duetsch2}.

We now wish to remove the cutoff. Formula (\ref{intfield})  will not, in general, make sense if we
straightforwardly attempt to take the limit $\theta \to 1$. Indeed if
$\theta$ could be set equal to 1 throughout the spacetime in
eq.~(\ref{intfield}), then the resulting formula for $\O(f)_{\eL_1(1)}$
would define an interacting field in the sense of Bogoliubov
\cite{bs}, with the property that the interacting field approaches the
free field in the asymptotic past. However, even in Minkowski
spacetime, it is far from clear that such an asymptotic limit of the
interacting field will exist (particularly for massless fields), and it
is much less likely that any such limit would exist in generic
globally hyperbolic curved spacetimes that do not become flat in the
asymptotic past.

To remove the cutoff in such a way that the limit does exist, we need
to know how the fields \eqref{intfield} change under
a change of the cutoff function $\theta$. If $\theta_1$ and
$\theta_2$ are two cutoff functions, each of which are 1 in an open
neighborhood of $V$ as above, then there exists a smooth function $h_-$
of compact support on $\M$ which is equal to $\theta_1 - \theta_2$ on the
causal past of the region $V$, and whose support does not intersect
the causal future of $V$. The unitary $U(\theta,\theta')$ defined by
\begin{equation}\label{Udef}
U(\theta_1,\theta_2) = S_{\eL_1(\theta_1)}(\eL_1(h_-))
\end{equation}
is then independent of the particular choice for $h_-$, and
one has \cite[thm. 8.6]{bf2}
\begin{equation}
\label{rel}
U(\theta_1, \theta_2) \ \O(f)_{\eL_1(\theta_1)} \ U(\theta_1, \theta_2)^{-1} =
\O(f)_{\eL_1(\theta_2)} \ ,
\end{equation}
for all fields $\O$ and all smooth scalar densities $f$
of compact support in $V$.

We now take a limit where the field remains fixed in
regions of increasing size in the interior of the spacetime, following 
the particular construction of~\cite{hw3} which is a realization of the idea of 
`adiabatic algebraic limit' that appeared first in~\cite{bf2}.
The construction makes use of the following
geometric fact (see lemma 3.1 of \cite{hw3}):
Let $(\M, g)$ be a globally hyperbolic spacetime. Then there exists a
sequence of compact sets, $\{K_n\}$, with the properties that (i) for
each $n$, $K_n \subset V_{n+1}$, where $V_{n+1} \equiv
{\rm int}(K_{n+1})$ (ii) $\cup_n K_n = \M$, and (iii) for each $n$,
$V_n$ is globally hyperbolic and $\Sigma \cap V_n$ is a Cauchy surface
for $V_n$, where $\Sigma$ is a Cauchy surface for $\M$.
For each $n$, let $\theta_n$ be a
smooth function with support contained in $K_{n+1}$ such that
$\theta_n = 1$ on an open neighborhood of $K_n$. Let $U_1 = \myid$ and let
$U_n = U(\theta_n, \theta_{n-1})$ for all $n>1$, where $U(\theta_n,
\theta_{n-1})$ was defined in eq.~(\ref{Udef}) above.  Our definition of the {\bf interacting field} is:
\begin{equation}\label{Ifield}
\O_{\rI}(f) \equiv \lim_{n \rightarrow \infty}   U_1
U_2  \dots U_n \, \O(f)_{\eL_1(\theta_n) } \, U_n{}^{-1} U_{n-1}^{-1} \dots U_1{}^{-1} \ .
\end{equation}
The existence of the limit (in the sense of formal power series) follows from the properties of the time-ordered products, see~\cite{hw3}.

The meaning of the sequence under the limit in eq.~\eqref{Ifield} for $n = 1, 2, \dots$, is easily understood as follows. Since
$U_1 = \myid$, the first element of this sequence is just the
Bogoluibov formula for this interacting field quantity with cutoff
function $\theta_1$. The second element of this sequence modifies the
Bogoliubov formula with cutoff function $\theta_2$ in such a way that,
according to eq.~(\ref{rel}) above, the modified Bogoliubov formula
with cutoff function $\theta_2$ agrees with the unmodified Bogoliubov
formula with cutoff function $\theta_1$ when the supports of all of
$f$ is contained within $K_1$. For the third element of 
sequence, the unitary map $U_3$ first modifies the Bogoliubov formula
with cutoff function $\theta_3$ so that it agrees in region $K_2$ with
the Bogoliubov formula with cutoff function $\theta_2$. The action of
the unitary $U_2$ then further modifies this expression so that it
agrees in region $K_2$ with the modified Bogoliubov formula of the
previous step. In this way, we have implemented the idea of ``keeping
the interacting field fixed in the interior of the spacetime'' as the
cutoff is removed. A particularly simple expression for the interacting field
$\phi_\rI(x)$ is hence obtained if $x \in K_1$, because then the unitaries
$U_n$ are absent in the formula. Unraveling the definitions, we find that, for the
case $\eL_1 = \phi^4$, the following
formula holds for $x \in K_1$
\ben\label{ifield}
\O_\rI(x) = \sum_{m,n \ge 0} \frac{(i\lambda)^{n+m}}{n!m!} \int \overline
\T_n\Big( \bigotimes_{j=1}^n \phi^4(y_j) \Big)
\T_{m+1}\Big( \O(x) \otimes \bigotimes_{k=n+1}^{n+m} \phi^4(y_k)  \Big) \, \prod_{j=1}^{n+m} \theta_1(y_j) dv_j \ .
\een
The interpretation of $\O_\rI$ as interacting fields is reinforced by the fact that the interacting field
equation
\ben
(\square_g - m^2) \phi_\rI^{}(x) = 4\lambda \phi^3_\rI(x)
\een
holds to all orders in $\lambda$, as a consequence of~\eqref{dyson}. 

As before (see \eqref{wlambda}), we denote by $\eW(\M, g)\rf{\lambda}$ the algebra of formal power series expressions in $\eW(\M, g)$. We define the {\bf interacting field algebra} $\cB_{\rI}(M, g)$ to be
the subalgebra of $\eW(\M, g)\rf{\lambda}$ generated by the interacting
fields, namely
\ben
\eB_\rI = {\rm alg} \{ \O_\rI(f) \mid \O \in \P(\M,g), \ \ f \in C^\infty_0(\M) \} \ .
\een
This definition of $\eB_{\rI}(\M, g)$ as a subalgebra of $\eW(\M, g)\rf{\lambda}$
depends on the choice of a family of compact sets, $K_n$, described above, as well as the choice of cutoff functions,
$\theta_n$. If we were to choose a different family, $\tilde{K}_n$, of
compact sets and a corresponding different family,
of cutoff functions, $\tilde{\theta}_n$, we will obtain a different subalgebra, $\tilde{\eB}_{\rI}(\M, g)$. However,
$\eB_{\rI}(\M, g)$ and $\tilde{\eB}_{\rI}(\M, g)$ can be shown to be isomorphic, and hence, in so far as the abstract
algebra of interacting fields is concerned, these choices are irrelevant.

In order to describe physical situations, one needs to construct concrete states on interacting field algebra $\eB_\rI$ representing such situations. Since, as we explained, 
the presentation of this algebra is such that fields are, in a sense, fixed by their ``initial values'' near some chosen Cauchy surface $\Sigma$ in the interior of the spacetime -- rather 
than in the asymptotic future or past -- states are naturally constructed by their ``initial values'' near $\Sigma$. As we have said, this is much more natural in curved spacetime. Building on such ideas, KMS states for the interacting field have for example been construced by~\cite{fred2}.  These constructions, performed for simplicity in flat space, are likely to have a straightforward 
extension e.g. to any static spacetime with compact Cauchy surface.

\vspace{.5cm}

An important and nontrivial feature of the interacting field algebra $\eB_{\rI}(\M,g)$ in any globally hyperbolic spacetime is its behavior under scaling of the spacetime metric. Classically, for any polynomial Lagrangian, $\eL$, we can define a scaling of the field and the coupling constants so that the action is invariant under scaling of the spacetime metric, $g \to \Lambda^2 g$. For example, for the Lagrangian
\ben
\eL = \half ( (\nabla \phi)^2 + m^2 \phi^2 + \xi R \phi^2) + \lambda \phi^4
\een
the associated classical scaling is $\phi \to \Lambda^{-1} \phi$, $m \to \Lambda^{-1} m$, $\xi \to \xi$, and 
$\lambda \to \lambda$. Any monomials in $\phi$ and its derivatives can similarly be assigned a scaling behavior. However, this classical scaling behavior cannot be maintained in the interacting quantum
field theory. Indeed, in our construction of composite fields and time-ordered products for the free field theory, we already noted that the classical scaling of these fields can be maintained only ``up to logarithmic corrections,'' thus implying a nontrivial behavior under scaling for the interacting fields.
Nevertheless, the following result can be proven \cite{hw3}: Let $\underline f$ be the collection of 
coupling parameters of the theory---such as, in our above example, $m^2, \xi$, and $\lambda$. Then there exists an 
isomorphism $\rho_\Lambda: \eB_{\rI}(\M, \Lambda^2 g, \underline{f}) \to \eB_{\rI}(\M, g, \underline{f}(\Lambda))$, 
which maps local fields to local fields. Thus, a change of scale (distances) can be compensated by a
corresponding change of the coupling constants, thereby 
leaving the theory unchanged. This change of the coupling constants, $\underline{f} \mapsto \underline{f}(\Lambda)$ is called the {\bf renormalization group flow}. The flow can be calculated order-by-order in perturbation theory. In our example theory, the renormalization group flow takes place in the 3-dimensional space of coupling constants $m^2, \xi, \lambda$. 
Theories such as this one where the flow takes place in a {\em finite dimensional} space of coupling constants are traditionally called ``renormalizable''. By contrast, theories where 
the flow necessarily takes place in an {\em infinite dimensional} space of coupling constants are called ``non-renormalizable''. 

In Minkowski spacetime, the dilations (i.e., the diffeomorphisms associated with a rescaling of global inertial coordinates) are conformal isometries with a constant conformal factor, so the renormalization group flow can be
reformulated in terms of the behavior of physical quantities, such as correlation functions of the vacuum state, under rescaling of coordinates (or, equivalently, rescaling of momenta). However, in a general curved spacetime, there
will not be any conformal isometries of any kind, so renormalization group flow must be defined in terms of behavior
under scaling of the spacetime metric, as described above. 
We emphasize that our definition of the renormalization group flow does not involve any preferred states 
(which do not exist in a general, curved spacetime), nor the idea of `cutoffs'. Cutoffs play an important role in a somewhat different concept 
of renormalization group flow in flat space often called ``Wilsonian RG-flow''. We do not see any canonical way of imposing a cutoff in 
general Lorentzian curved spacetime, and furthermore, any cutoff is in fundamental conflict with the idea of defining the theory 
in a ``local and covariant manner''. Further discussion of the relationship between various different formulations of renormalization group flow can be found in~\cite{bdf}.


\subsection{Yang-Mills fields}
\label{sec4}

For simplicity, up to this point we have restricted our discussion QFTCS to scalar fields. However, there are other types of fields that occur in nature\footnote{Indeed, the existence of a scalar field in nature has only recently been confirmed by the discovery of the Higgs particle.}, namely spinor and gauge (Yang-Mills) fields. To treat spinor fields, one must replace the canonical commutation algebra $\eA$ of section 2.1 by a corresponding ``canonical anti-commutation algebra''. However, aside from this important change, spinor fields can be dealt
with without introducing any major new conceptual innovations. In particular, analogs of $\eW$, $\eB_0$, and $\eB_\rI$ can be defined for spinor fields in close parallel with the scalar case; see e.g.~\cite{extended} for details.

However, significant conceptual innovations are needed for the construction of the quantum field theory of gauge fields. All of the difficulties are associated with the fact that the physical observables in a gauge theory are not the gauge fields themselves but equivalence classes of these fields under gauge transformations. In particular, the classical equations of motion for gauge fields are not deterministic---they do not determine the gauge---so one does not have unique advanced and retarded propagators. One obvious idea would be to fix the gauge, thereby making the classical dynamics deterministic. However, the gauge cannot be (completely) fixed in a local and covariant manner, thereby making it extremely difficult to impose the requirement that the theory be local and covariant\footnote{The corresponding difficulty for gauge fields in Minkowski spacetime is that the gauge cannot be completely fixed in a Poincare invariant manner.}. An alternative, relatively straightforward, strategy can be employed for the case of a free (i.e., Maxwell) gauge field, $A_\mu$, to construct a quantum field theory at the level of the algebra $\eA$: Instead of smearing $A_\mu$ with arbitrary test vector fields $f^\mu$, one defines $A_\mu (f^\mu)$ only for $f^\mu$ that satisfy
\ben
\nabla_\mu f^\mu = 0 \, .
\label{divf}
\een
Such observables are invariant under the gauge transformation $\delta A = d\chi$, since
$\int \nabla_\mu \chi f^\mu = - \int \chi \nabla_\mu f^\mu = 0$. Indeed, the restriction to smearing with such $f^\mu$ is equivalent to working with the Maxwell field tensor $F_{\mu \nu} = 2 \nabla_{[\mu} A_{\nu]}$ as the fundamental observable. This yields a satisfactory theory at the level of the algebra $\eA$, see e.g.~\cite{fewsterpf,sanders3,finster} for details. 
But it is far from clear how to construct an analog of $\eB_\rI$ for models with interaction, such as Yang-Mills theory, by proceeding this manner.

Instead, we shall we shall proceed by an elegant trick for constructing gauge invariant observables for Yang-Mills theory by
introducing additional fields---the ``Fadeev-Popov ghost fields,'' $c$ and $\overline c$, together with an ``auxiliary field'' $b$---and then eliminating the extra degrees of via the imposition of a
BRST-symmetry condition. The observables of the resulting theory are equivalent to the gauge invariant observables
of the original Yang-Mills theory. As has been shown in~\cite{h3},
one can implement the procedure consistently obtaining in the end an algebra of gauge invariant quantum observables $\eB_\rI$
in perturbative interacting Yang-Mills theory.

We now very briefly outline this construction. The dynamical field in Yang-Mills theory is a connection, $D_\mu$, on a principal fiber bundle with Lie group $G$. For simplicity, 
we consider the trivial bundle $\M \times G$, so that we may write
\ben
D_\mu=\nabla_\mu + i\lambda A_\mu
\een
where $A_\mu$ is a one-form field on $\M$ valued in the Lie algebra, $\frak{g}$, of $G$. We have 
introducted a coupling constant $\lambda$ in anticipation of the fact that we will later perform 
a perturbative expansion of the theory. 
The field strength $F_{\mu\nu}$ is the $\frak{g}$-valued two form given by
\ben
\frac{i}{\lambda} F = dA + i\lambda[A,A]
\een
The Yang-Mills Lagrangian is given by
\ben
\eL = \frac{1}{4} \langle F_{\mu\nu},F^{\mu\nu} \rangle \, ,
\label{yml}
\een
where $\langle \ . \ , \ . \ \rangle$ denotes the Cartan-Killing form of $\frak{g}$, and we assume that $G$ is compact and semi-simple, so that this Cartan-Killing form is positive-definite. This Lagrangian is invariant under the infinitiesimal  gauge transformations
\ben\label{gauge}
\delta A = d \chi - i\lambda [\chi, A] = D \chi \ , 
\een
where $\chi$ is an arbitrary $\frak{g}$-valued scalar field on $\M$.

We now introduce additional $\frak{g}$-valued scalar fields $c, \overline c$, and $b$. The fields $c,\bar c$ are independent and are Grassmann-valued (as well as $\frak{g}$-valued). We define an infinitesimal ``BRST-transformation,'' $\Q$, acting on the fields $(A,c,\overline c, b)$ by 
\ben
\Q A = Dc \ , \quad
\Q c = -\half i\lambda [c,c] \ , \quad
\Q \bar c = b \ , \quad
\Q b = 0 \ .
\een
Note that the transformation law for $c$ is exactly analogous to the gauge transformation~\eqref{gauge}, i.e. $c$ may be 
thought of as a Grassmann-valued counterpart of the gauge transformation function $\chi$. The Grassmann-property (and Jacobi-identity) are 
used to show that  $\Q$ is nil-potent, $\Q^2 = 0$. Note that $\Q$ changes the 
``Grassmann-grading'', i.e. maps Grassmann-even expressions to Grassmann-odd expressions, and vice versa. 
We now modify the Yang-Mills Lagrangian \eqref{yml} to
\ben
\tilde \eL = \eL + \Q \Psi
\een
where the density $\Psi$ is given by
\ben
\Psi = \langle \nabla^\mu A_\mu, \bar c \rangle - \half t \langle b,\bar c \rangle
\een
with $t \in (0,1]$. The choice $t=1$ is called the ``Feynman gauge,'' while the limit $t \to 0$ is called the ``Landau gauge.'' Note that the new Lagrangian $\tilde\eL$ is BRST-closed, i.e. $\Q \tilde \eL = 0$: 
Indeed, $\Q \eL = 0$ because the action of $\Q$ on expressions involving only $A$ corresponds exactly to a gauge transformation~\eqref{gauge}, whereas $\Q(\Q \Psi) = 0$ because $\Q$ is nil-potent. 
A key difference between $\tilde \eL$ and $\eL$ is that Euler-Lagrange equations arising from
$\tilde \eL$ are hyperbolic for $A,c$, and $\bar c$ and are algebraic in $b$, so the classical dynamics defined by $\tilde \eL$ is deterministic. Of course, the theory defined by $\tilde \eL$ is a different theory, and we are interested in the theory defined by $\eL$. Nevertheless, there is a direct relationship between gauge invariant observables in the original Yang-Mills theory defined by $\eL$ and observables in the new theory defined by
$\tilde \eL$: The gauge invariant polynomials  $\O = f(D^{k_1} F, \dots , D^{k_n}F)$ (with $f$ an invariant multi-linear form on $\frak{g}$) 
are BRST-invariant since on such observables, $\Q$ acts just like a gauge-transformation. 
Futhermore, modulo BRST-exact observables, these are {\em all} observables with vanishing ``ghost number'' in the new theory, see e.g.~\cite{henneaux}. 
This suggests the following strategy to define the quantum theory of a Yang-Mills field: First define the algebra of interacting fields $\tilde \eB_\rI$ of the
theory corresponding to $\tilde \eL$. This should be possible, since the Euler-Lagrange equations
for $\tilde \eL$ do not suffer from the gauge indeterminism of the theory defined by $\eL$. Then define a
quantum version of $\Q$ on $\tilde \eB_\rI$ which acts as a nilpotent, graded, derivation. The cohomology
(closed modulo exact elements) at ghost number zero should naturally be viewed as the algebra of gauge invariant polynomial observables of the {\em original} theory.

The difficult step is to define a quantum version of $\Q$. The point is that such an object must be defined so as to
be compatible with renormalization~\cite{h3,fred3}. Informally speaking, one would like to define $\Q$ as the (graded) commutator 
with a ``BRST-charge'', which itself is obtained from a corresponding conserved Noether current associated with the BRST-invariance of $\tilde \eL$, defined in turn 
as an interacting field in the algebra $\tilde \eB_\rI$ in the manner described in the previous section. In order to arrive at a consistent
definition of the corresponding BRST charge-``operator'', this current should be conserved (as a quantum operator), but it is not obvious from the outset that this can actually be achieved. In fact, 
an arbitrary renormalization prescription of the kind described above will not lead to a conserved BRST-current. This failure to be conserved is called an 
``anomaly''. By deriving suitable ``consistency conditions''\footnote{A similar consistency condition 
was derived in a somewhat different formalism in~\cite{reijzner}.} on this anomaly, one can show~\cite{h3} that an arbitrary renormalization prescription, consistent with 
all of the properties listed in thm.~\ref{Tuniqueness}, can always be modified so as to remove this anomaly\footnote{This statement is valid for 
pure Yang-Mills theory. For Yang-Mills theories with matter fields in non-real representations of the gauge group, this need not hold in general. Such 
theories simply cannot be constructed consistently at the quantum level, at least not via the BRST-method.}.

\section{Open issues}
\label{sec5}

In this final section, we briefly discuss two fundamental outstanding issues related to QFTCS. In subsection~\ref{sec:ope} we discuss a possible approach to the formulation of interacting QFTCS via operator product expansions. In subsection~\ref{sec:qg}, we consider the formulation of quantum gravity from the viewpoint of QFTCS.

\subsection{Nonperturbative formulation of interacting QFTCS}\label{sec:ope}

The perturbative construction of interacting QFTCS given in the previous section attempts to represent the interacting quantum field as an element of the algebra $\eW$ of the free quantum field theory. While this can be done to any finite order, the algebra of interacting fields cannot plausibly be isomorphic to a subalgebra of $\eW$, simply because $\eW$ by definition consists of
elements that are, in a sense, of finite ``polynomial order'' in the free field, whereas the order of the terms in the perturbation series clearly grows unboundedly. Thus, in order to even ask the
question about convergence of the series, one would have to go to a suitably enlarged algebra and it is not clear to us what that object might be. The situation is even worse if one is calculating properties of special states, such as ground states. Even if the interacting fields were elements of the same algebra $\eW$ for all $\lambda$ and had a common set of states, as Dyson noted more than 60 years ago, for a theory such as the one with interaction Lagrangian $\lambda \phi^4$, a ground state $\omega_0 (\lambda)$ cannot be expected to be analytic in $\lambda$ at $\lambda = 0$, since no ground state can exist when $\lambda < 0$. Thus, perturbative expressions for quantities such as the $S$-matrix should not
converge---even if it were the case that notions of ``in'' and ``out'' particle states could be defined. Thus, the perturbative approach described in the previous section would appear to be a very poor way to formulate QFTCS for interacting fields.

What we would really like to do, of course, is {\em directly} write down the algebra, $\eB_\rI$, of local field observables for the interacting field, in analogy to our construction of the algebra $\eB_0 \subset \eW$ for the free field. However, our ability to define the algebra $\eB_0$ traces back to the explicit product formula \eqref{wick}, which allowed us to express the product of any two
elements of $\eW$ as a {\it finite sum} of other elements of $\eW$. This property, in turn, traces back to the fact that
the extremely simple commutation relations satisfied by $\phi_0(f)$, namely $[\phi_0(f_1), \phi_0(f_2)] \propto \myid$. We cannot expect that any simple product relations of this sort will hold for an interacting quantum field. It therefore would
seem a hopeless task to directly construct an interacting field algebra $\eB_\rI$.

Nevertheless, we believe that a possible route towards a nonperturbative formulation of QFTCS for interacting fields
arises from the fact that it appears likely that interacting quantum fields satisfy sufficiently simple ``product relations'' in the limit of small separation of points. More precisely, it is known that order-by-order in perturbation theory, interacting
quantum fields satisfy relations known as {\it operator product expansions} (OPEs).
What is meant by this is the following. Let us first introduce a ``basis'' of composite fields $\O_A$. The label incorporates both the kind of field
(such as $\phi^2, \phi^6, ...$ in the case of a free KG-field), as well as any tensor/spinor indices (such as in $\phi \nabla_{(\mu} \nabla_\nu \nabla_{\sigma)} \phi$). The OPE
states that the product of any number of composite fields can be expanded in an asymptotic series of the form
\ben
\O_{A_1}(x_1) \cdots \O_{A_n}(x_n) \sim \sum_{C} \mathcal{C}_{A_1 ... A_n}^C(x_1, ..., x_n) \O_C(x_n) \ .
\een
The coefficients $\cC$ appearing in the expansion are $\CC$-valued distributions that should be locally and covariantly in terms of the spacetime
geometry $(\M,g)$, and which also, of course, depend on the particular theory under consideration. They are analogous in a rough sense to the structure constants of
a finite dimensional algebra. The asymptotically equal sign $\sim$ expresses that, if both sides of the expansion are inserted into a suitably well-behaved state
$\omega$ (a Hadamard state for the case of a free KG-field), and if the sum over $C$ is carried out to operators of a sufficiently high dimension, then the
right side should approximate the left side well for short distances, $x_1, ..., x_{n-1} \to x_n$. For the free KG-field, the OPE coefficients can be found concretely
in terms of the Hadamard parametrix, $H$. For example, for the simple case $\O_A=\O_B=\phi, \O_C=\myid$, we have $\cC_{AB}^C(x_1,x_2) = H(x_1,x_2)$.
In the context of a perturbative interacting theory in curved space, an algorithm for finding the coefficients was given in~\cite{h4}. Remarkably, for Euclidean perturbative $\phi^4$-theory,
it has been shown that the series on the right side actually {\em converges}, at least at arbitrary but fixed order in perturbation theory~\cite{hoko,koke}. If, as we expect, this also
were found to be true at spacelike related points for theories on Lorentzian curved spacetimes, then this would
strongly reinforce the view that the OPE coefficients contain the entire local information about the theory, even on a curved spacetime.

The OPE-coefficients are expected to satisfy a kind of ``factorization rule'' in a situation in which a subset of points, say $(x_1, ..., x_m)$ is much ``closer to each other than the rest''. In that situation, a relation of the type
\ben
\cC_{A_1 ... A_n}^C(x_1, ..., x_n) \sim \sum_B \cC_{A_1 ... A_m}^B(x_1, ..., x_m) \cC_{BA_{m+1} ... A_n}^C(x_m, ..., x_n) \ .
\label{assoc}
\een
should hold. 
This factorization rule has been shown to hold in Euclidean $\phi^4$-theory in~\cite{h5} to arbitrary but finite order in perturbation theory; again the sum on the right side is shown to actually converge, and ``$\sim$'' can actually be replaced by an equality sign! Eq.~\eqref{assoc} can be thought of as a version of the associativity law for the quantum fields, holding at the level of the coefficients $\cC$. 

A possible approach towards a formulation of QFTCS for interacting fields would be to define the theory via its collection of
OPE coefficients. The OPE coefficients would be required to satisfy the factorization rule \eqref{assoc} together with a list of other natural properties; see \cite{hw5} for further discussion. The basic philosophy of this approach is that the OPEs would then fully determine the quantum field theory in a local manner, even though one may not easily
be able to construct the field algebra and the state space from the OPEs. This would be closely analogous
to the situation in classical field theory, where the field equations uniquely determine the theory in a local manner, but one may not be able to easily construct the space of all solutions from the field equations. The advantage of formulating
QFTCS in terms of OPEs is that to define OPEs, one must specify a collection of ``$\CC$-number'' distributions. These
can be specified without reference to a ``background field algebra'' or ``background states.''  In particular, there any no
obvious reasons why the OPE coefficients of an interacting theory could not be constructed perturbatively, with a convergent perturbation series.

Unfortunately, it seems to be extremely difficult to find non-trivial solutions to the associativity law from first principles rather than by indirect methods that rely on perturbation theory.  However, there is some hope that a self-consistent equation for the OPE-coefficients may be obtained. The idea is that interacting quantum field theories are normally
labeled by some coupling parameter, such as $\lambda$ in $\lambda \phi^4$-theory. The OPE-coefficients thus have to be functions of $\lambda$. In~\cite{h6}, it is shown that, again
in perturbation theory, there exists a relation of the type
\ben\label{eq:thmint1}
\begin{split}
& \frac{\partial}{\partial \lambda} \, \cC_{A_{1}\dots A_{N}}^{B}(x_{1},\dots, x_{N})=  \\
& - \int_y  \bigg( \cC_{\O A_{1}\dots A_{N} }^{B}(y,x_{1},\ldots,x_{N})-\sum_{i=1}^{N}\sum_{[C]\le [A_{i}]} \!\!\!\cC_{\O A_{i}}^{C}(y,x_{i})\, \cC_{A_{1}\dots \widehat{A_{i}}\,C\dots A_{N}}^{B}(x_{1},\dots, x_{N}) \\
&\qquad\qquad\qquad \qquad\qquad\qquad\qquad -\sum_{[C]< [B]} \cC_{A_{1}\dots A_{N}}^{C}(x_{1},\ldots,x_{N})\, \cC_{\O C }^{B}(y,x_{N}) \bigg) \, ,
\end{split}
\een
where $\widehat{A_{i}}$ denotes omission of the corresponding index, where $[A]$ indicates the dimension of the field $\O_A$,
and where the operator $\O$ ($= \phi^4$ in this example) is the interaction. The spacetime integral over $y$ can be shown to converge absolutely. Although it is derived within perturbation theory (i.e. in the sense
of formal power series in $\lambda$), the final form of~\eqref{eq:thmint1} no longer makes any reference to perturbation
theory and thus may be viewed as a self-consistent relationship between the OPE-coefficients. Eq.~\eqref{eq:thmint1} is a first order differential equation in $\lambda$, and since $\lambda=0$
corresponds to the free field theory, the ``initial values'' of all OPE-coefficients are known at $\lambda=0$. Thus, it is conceivable that one could actually show that a unique solution to~\eqref{eq:thmint1} must exist (beyond the level of formal power series). We view this as a promising approach to a non-perturbative definition of the OPE-coefficients.

In summary, and as discussed further in \cite{hw5}, we believe that OPEs provide a promising approach toward the formulation
of QFTCS that is independent of, and different from, perturbative methods.

\subsection{Quantum gravity} \label{sec:qg}

As we have seen in this review, the formulation of QFTCS assumes the presence of
a {\it classical} spacetime metric $g$. Consequently, QFTCS can, at best, be an approximation to a more fundamental
theory of ``quantum gravity,'' wherein the metric is no longer treated as a classical field. We shall not attempt here to review---or even mention---the various approaches that have been taken towards the formulation
of a quantum theory
of gravity, their successes, and their difficulties. Suffice it to say that no approach has yet been formulated that has
the conceptual clarity and mathematical rigor of QFTCS. The question we do wish to briefly address here is whether QFTCS
provides any insights towards the formulation and properties of quantum gravity.

One can attempt to formulate a quantum theory of gravity by expanding the metric $g$ about a background
solution $g_0$, i.e., by writing
\ben
g = g_0 + \gamma \, .
\label{g0}
\een
The idea is now to treat $\gamma$ as a quantum field propagating in the classical spacetime $(\M, g_0)$, and to formulate a quantum theory of $\gamma$ by the procedures of QFTCS. However, in order to be in accord with
the fundamental principle of general relativity that no ``background structure'' appear in the theory apart from the spacetime manifold $\M$, one would like the final theory to be independent of the choice of $g_0$ appearing in the
expansion \eqref{g0}. To impose this requirement, $\gamma$ should satisfy a complicated ``gauge symmetry'' that can be obtained by applying an arbitrary diffeomorphism $\psi$ to $g$ and then re-expanding $\psi^* g$ about $g_0$. Indeed, even
if one does not care about ``fundamental principles'' one would need to impose a gauge symmetry of this sort
in order to have deterministic dynamics, since Einstein's equation does not determine $g$ but only the diffeomorphism
equivalence class of $g$. Thus, one cannot get well defined ``propagators'' for $\gamma$ without making a gauge
choice, and it would be natural to insist that the resulting theory does not depend upon the choice of gauge.

In this regard, the situation with regard to formulating QFTCS for $\gamma$ may appear to be similar to the
situation for formulating QFTCS for a Yang-Mills field, $A$, which has a somewhat similar-looking gauge symmetry. For the Yang-Mills field, an appropriate route toward formulating QFTCS was to formulate
the theory in terms of gauge invariant local observables, which can be done via the BRST procedure discussed
in subsection~\ref{sec4} above. However, it is clear that such a route cannot work in the case of quantum gravity, simply because the nature of diffeomorphism gauge symmetry is such that {\it there do not exist any gauge invariant local observables}.
Diffeomorphisms do not merely ``transform fields at a point'' but they ``move the points themselves around,'' so there cannot be any physical meaning to the properties of the metric ``at a given spacetime location.'' Thus, one cannot
hope to formulate QFTCS for $\gamma$ in terms of gauge invariant local observables\footnote{However, this does not preclude the possibility of using key ideas from QFTCS in the formulation of quantum gravity; see~\cite{bfr} for recent progress in this regard.}. This difficulty is closely related to the ``problem of time'' encountered in approaches to quantum gravity that attempt to provide the quantum state of the metric at a given ``time.'' If the theory is to respect the diffeomorphism gauge symmetry, the state cannot depend upon
``time,'' thereby making its interpretation extremely nebulous.

Although, as we have just argued, the above approach cannot yield a quantum theory of gravity, we can
successfully obtain a theory of {\it linearized gravity} off of an arbitrary background solution $g_0$
by following the analog of the procedure for Maxwell fields discussed near eq.~\eqref{divf} above. For linearized gravity,
the gauge transformations of $\gamma$ are simply $\gamma \to \gamma + \pounds_\xi g_0$, where $\xi$ is an
arbitrary vector field on $\M$. One can then construct an analog of the algebra $\eA$ of section 2.1
for the quantum field $\gamma_{\mu \nu}$ by defining its ``smearing'' only with symmetric
test tensor fields $f^{\mu \nu}$ that satisfy\footnote{In Minkowski spacetime, a symmetric tensor field $f^{\mu \nu}$ satisfying
\eqref{fgauge} can be written in the form $f^{\mu \nu} = \partial_\alpha \partial_\beta U^{\alpha \mu \beta \nu}$, where
$U^{\alpha \mu \beta \nu}$ is antisymmetric in its first and last pairs of indices and also satisfies $U^{\alpha \mu \beta \nu} = U^{\beta \nu \alpha \mu}$. From this and the linearized field equation, it can be seen that the smeared observables $\gamma(f)$ are equivalent to using the smeared linearized Weyl tensor as the fundamental observables
of the theory. However, if perturbing off of a nontrivial background, the gauge invariant observables $\gamma(f)$ with $f$ satisfying
\eqref{fgauge} are no longer equivalent to the linearized Weyl tensor (which, itself, is no longer gauge invariant).}
\ben
\nabla_\mu f^{\mu \nu} = 0 \, .
\label{fgauge}
\een
Note that the resulting smeared fields $\gamma(f)$ are gauge invariant since
\ben
\int_\M (\pounds_\xi g_0)_{\mu \nu} f^{\mu \nu} = 2 \int_\M \nabla_{(\mu} \xi_{\nu)} f^{\mu \nu} =
- 2 \int_\M \xi_\nu \nabla_\mu f^{\mu \nu} = 0 \, .
\een
The classical advanced and retarded are well defined when smeared with test fields satisfying \eqref{fgauge},
and this is all that is needed to construct the analog of $\eA$ for linearized gravity; for details see~\cite{fewster2}. Consequently, all of the 
constructions and analyses
given in section 2 for the free scalar field can be repeated, in parallel, for the linearized gravitational field, including the Hawking
effect and the behavior of cosmological perturbations\footnote{To treat cosmological perturbations, one must linearize all fields off of a classical background Einstein-matter solution.}. Thus, despite the lack of a quantum theory of gravity,
one can rather confidently make predictions about the ``radiation of gravitons'' by a black hole and the generation of
``tensor perturbations'' by quantum field effects in cosmology.

Nevertheless, if is far from clear whether an analog of $\eW$ can be defined for the linearized gravitational field,
and it is far from clear that perturbative QFTCS can be defined for $\gamma$---even at second order\footnote{If one
has appropriate ``fixed asymptotic regions'' of the spacetime, then it should be possible to define an $S$-matrix
to all orders in perturbation theory, since such a quantity would be ``gauge invariant'' in an appropriate sense. However,
as we have argued in the introduction, the quantities of main interest in gravitational physics are local observables, not relations between ``in'' and ``out'' states (even when such notions can be defined). Furthermore, it is worth mentioning that, as is well known, on account of the ``non-renormalizability'' of the gravitational action, the determination of $S$-matrix to all orders will depend upon an infinite number of parameters of the renormalization scheme, and thus has relatively little predictive power.}. Furthermore, even if a perturbative QFTCS could somehow be defined for $\gamma$,
it is very difficult to see how it could give rise to a theory that makes predictions about local observables in a
manner that is compatible with the basic principles of general relativity.

Thus, it seems clear that ideas that go beyond the principles of
QFTCS as described in this review will be needed to formulate a quantum theory of gravity. Nevertheless,
as a consequence of its prediction of such phenomena as the Hawking effect,
QFTCS has already provided some remarkable insights into the nature of quantum gravity, and we believe that
it will continue to do so in the future.

\vspace{1cm}
\noindent
{\bf Acknowledgements:} The research of S.H. was supported in part by ERC grant QC \& C 259562. 
The research of R.M.W. was supported in part by NSF grant PHY 12-02718 to the University of Chicago.

\appendix

\section{Distributions, scaling degree, and wave front sets}
\label{appA}

The objects appearing in quantum field theory such as $n$-point functions, time-ordered
products, etc. are singular and therefore best viewed as distributions. A distribution $u$ on a $d$-dimensional manifold $X$ is a complex linear functional $u: C^\infty_0(X) \to \CC$ for which
there is a constant  $c_K$ and an $N \in {\mathbb N}_0$ for each compact $K \subset X$ such that
\ben
|u(f)| \le c_K \sum_{k \le N} \sup_{x \in K} |D^k f(x)|
\een
for any $f \in C^\infty_0$ having support within $K$, where $D$ is any derivative operator on $X$. For instance, the delta ``function'' on $\RR$ concentrated at $0$, which is defined by $\delta(f) = f(0)$, evidently satisfies the above estimate with $N=0$. The $n$-th derivative $\delta^{(n)}$ satisfies the criterion with $N=n$. Also, any smooth function $u$ defines a distribution
via $u(f) = \int u(x) f(x) dv$ for a given integration element $dv$ on $X$. Such a distribution is called smooth, and more generally, a distribution is called smooth at $x_0 \in X$ if it can be represented in that way for $f$ having support sufficiently close to $x_0$. The complement of the set of all such $x_0$ is called the ``singular support''
${\rm singsupp} (u) \subset X$. The notion of singular support is not very informative since it gives no insight into the precise nature of the singularity at a given $x_0 \in
{\rm singsupp}(u)$. This shortcoming can be dealt with by introducing more refined concepts
to characterize singularities. Two such concepts of particular relevance for QFTCS are
that of the {\bf scaling degree} and that of the {\bf wave front set}.

The {\bf scaling degree} of a distribution $u$ at a point $x \in X$ basically describes the ``degree of
divergence'' at $x$, if any. It is defined more formally as follows. Choose an arbitrary
chart $(U,\psi)$ near $x$ and let $u_\psi(f) = u(f \circ \psi^{-1})$ the pull back
of $u$ to $\RR^d$, defined for $f$ supported in $\psi[U]$. Without loss of generality
we may assume that $\psi(x)=0$, and we define $f_\epsilon(y) = \epsilon^{-d}f(y/\epsilon)$.
The scaling degree is given by
\ben
{\rm sd}_x(u) = \inf \{ \delta \in \RR \mid \lim_{\epsilon \to 0^+} \epsilon^\delta u_\psi(f_\epsilon) = 0 \ \
\text{for all $f$ supported in $\psi[U]$} \} \ .
\een
It is easily checked that the definition is independent of the choice of chart $(U,\psi)$. For 
example, the scaling degree of the distributions $(x+i0^+)^{-n}$ on $\RR$ at $x=0$ is $n$,
whereas it is $0$ at any other point $x \neq 0$. The scaling degree of the $n$-th derivative
$\delta^{(n)}$ of the delta distribution on $\RR$ at $x=0$ is  likewise $n$ whereas it is $-\infty$ for
$x \neq 0$. The scaling degree ``doesn't see
logarithms'': The scaling degree of $\log^n (x+ i0^+)$ is $=0$ for any $n$ at any point $x \in \RR$.

The concept of {\bf wave front set}~\cite{hor} does not characterize the strength of a singularity, but
rather its nature from the point of view of momentum space.
To define the wave front set, assume first a distribution $u$ of compact support contained
in some chart $(U,\psi)$ of $X$. Then we may define the Fourier transform in that
chart by $\hat u_\psi(k) = u[\exp i\psi( \ . \ ) \cdot k]$. If $u$ is smooth within
$U$, then it is easy to see that there holds
\ben\label{estim}
|\hat u_\psi(k)| \le c_N (1+|k|)^{-N} \qquad \text{for all $N \in \mathbb N$,}
\een
for some constants $c_N$. For a general distribution supported in
$U$, we say that $k_0 \neq 0$ is a {\em singular direction} if there is no open cone $\Gamma$ around $k_0$ such that 
eq.~\eqref{estim} holds uniformly in $\Gamma$. If $x_0 \in X$ and if $u$ is an arbitrary
distribution, we say that $(\psi(x_0), k_0)$ is in the wave front set $\WF(u_\psi)$ of
$u_\psi$ if $k_0$ is a singular direction for $\chi u$ for all cutoff functions
$\chi$ supported in $U$ such that $\chi(x_0) \neq 0$. The wave front set of $u_\psi$
is a subset of $\psi[U] \times (\RR^d \setminus\{0\})$. The pull-back
\ben
\WF(u) = \bigcup_{{\rm charts} \ \ (U,\psi)} (\psi^{-1})^* \WF(u_\psi) \subset T^* X \setminus 0 \ ,
\een
can be shown to be invariantly defined (i.e. independent of the choice of atlas of $X$
for the given differentiable structure), and is simply called the ``wave front set''.

The notion of wave front set is applied above in eq.~\eqref{hadacon} to characterize
Hadamard 2-point functions $W_{2}$ ($X = \M \times \M$ in that example) of the free KG field. It can also be used to characterize the wave front set of an $n$-fold time ordered product $\T_n$ ($X= \M \times ... \times \M$~\cite{bf1,bf2} ($n$ copies) in that case), or of the $n$-point functions of $n$ interacting fields or their OPE coefficients~\cite{h5}.

One of the most important
uses of wave-front sets in QFTCS is to characterize situations in which
the product of distributions is defined. In fact, the following theorem holds:
Let $u,v$ be distributions on $X$. If $\WF(u) + \WF(v)$ (element-wise addition) does
not contain a zero cotangent vector in $T^*X$, then the distributional product $uv$ is
naturally\footnote{Note that there may be other exotic ways to
define the product of distributions even if the wave-front criterion is not fulfilled; the key point here is that the so-defined product is continuous
in some natural topology of distributions, i.e. when $u,v$ are approximated by smooth functions in an appropriate way.} 
defined. More generally, for a set of $n$ distributions, if $\sum_j \WF(u_j)$
does not contain a zero cotangent vector, then $\prod_j u_j$ is defined.

As an example, consider the distribution $(x + i0^+)^{-1}$, whose
wave front set is found to be $\{ (0,k) \mid k > 0 \}$.
The square -- and in fact any power -- is therefore well defined. Next, consider $\delta(x)$, which
has wave front set $\{(0,k) \mid k \neq 0 \}$. Its square is therefore not
well defined. One way to think about these examples is that in the first case,
$(x+i0^+)^{-1}$ is, by definition, the boundary value of an analytic function.
Whence its powers also are the boundary value of an analytic function, and
hence automatically defined. By contrast, the distribution $\delta(x)$ is
not the boundary value of an analytic function, whence its square is
not automatically defined. More generally, the relationship between distributional
boundary values and the wave-front set is that if $u(x+iy)$ is an analytic
function in $U \times \Gamma$, where $U \subset \RR^d$ and $\Gamma$ is some open cone
having finite scaling degree in $y$ at $y=0$ at $x \in U$ uniformly in $\Gamma$, then the wave front
set of the distributional boundary value $u(x) = \lim_{y \in \Gamma, y \to 0} u(x+iy)$
is contained in
\ben\label{dualcrit}
\WF(u) \subset U \times \Gamma^* \ ,
\een
where $\Gamma^* = \{ k \in (\RR^d)^* \setminus 0 \mid \langle k, y \rangle \le 0 \ \forall y \in \Gamma\}$
is the dual cone. This criterion can also be applied in any (analytic) manifold $X$
by localizing $u$ in a chart $(\psi,U)$. 

This relationship between wave front set and distributional boundary values is relevant 
in QFTCS, because many distributions involve some ``$i0^+$-prescription''. 
The wave front set of the two-point function $W_2$ of a Hadamard state for instance can be determined from~\eqref{localhada} and~\eqref{dualcrit}, since the $i0^+$-prescription effectively states that
$W_2$ is given in a sufficiently small open set $U \subset \M \times \M$ by a distributional boundary value (with cone locally given by $\Gamma = \cup_{(x_1,x_2) \in U} V^+_{x_1} \times V^-_{x_2}$). This can be used to deduce the wave front condition~\eqref{hadacon}. 

Two important applications of the above product criterion for distributions are the following.
Consider first the $k$ times contracted Wick-product eq.~\eqref{wick1} entering in the 
definition of the algebra $\eW$. The right side involves the product of the distributions $F_n,
G_m, W_2$. The first two distributions satisfy the wave front condition~\eqref{wfw}, whereas
$W_2$ satisfies~\eqref{hadacon}. It is easily seen that the product criterion is satisfied, 
whence the $k$ times contracted product is indeed well-defined. 

As the second example, consider the  time-ordered 2-point function
$W_2^\T(x_1,x_2) = \omega(\T_2(\phi(x_1) \otimes \phi(x_2))) $
associated with a Hadamard state $\omega$, also called a ``Feynman propagator''.
The wave front set of the time-ordered product is e.g. found to be
\ben
\begin{split}
\WF(W_2^\T) =& \{ (x_1, k_1; x_2, k_2) \in T^* \M^2 \setminus 0 \mid k_{1/2} \in V^\mp \ {\rm if} \ x_{1/2} \in J^+(x_{2/1}), k_1 \sim -k_2 \} \\
&
\cup \{ (x_1,k_1; x_2,k_2) \in T^* \M^2 \setminus 0 \mid k_1 = - k_2, x_1 = x_2 \} \ .
\end{split}
\een
Because $\WF(W_2^\T)$ contains e.g. the point $(x,k;x,-k) $ for any $k \neq 0$, 
$\WF(W_2^\T) + \WF(W_2^\T)$ contains the zero co-vector. The product criterion is not 
fulfilled, and thus the square of $W_2^\T$ cannot straightforwardly be defined as a distribution. This problem shows up precisely when one naively tries to define
the product via Fourier transform e.g. for the vacuum state in Minkowski space,
and is directly related to the logarithmic divergence of the ``fish-graph'' in
Feynman diagram language. However, for $x_1 \neq x_2$ the criterion {\em is} fulfilled and
$[W_2^\T(x_1,x_2)]^2$ {\em can} be defined for such points. Thus, the ``renormalization'' required to define the time ordered product
$\omega(\T_2(\phi^2(x_1) \otimes \phi^2(x_2)))$ 
corresponds precisely to obtaining an {\em extension} of 
this distribution to the ``diagonal'' of $\M \times \M$, i.e.
in some sense, no problems arise other than for coincident points. This is a rather 
non-trivial point in curved spacetime, because the behavior of null-geodesics (points
where the ``propagators'' are singular) can be very different from flat spacetime.  

These considerations can be generalized to the construction of higher 
order time ordered products. For instance, 
in order to define the expectation value $\omega(\T_n(\otimes_j \phi^{k_j}(x_j)))$ in a Gaussian, Hadamard state for mutually distinct points $x_j \in \M$, one may apply the 
Wick-product and ``causal factorization'' formulas. This leads to an expression in terms of
a product of Feynman propagators $W^\T_2(x_i, x_j)$, where $e=(ij)$ run through 
the edges of an abstract Feynman 
graph $\eG$ with incidence number of the $j$-th vertex $\le k_j$.
To the product, we may again apply our criterion and conclude that it exists 
away from all ``diagonals'', i.e. for the open subset of $\M^n$ of points such 
that $x_i \neq x_j$ for all $i \neq j$.  
Again, the important point is that the ``extension'' has to be performed only on 
the ``small'' subset of diagonals in $\M^n$, and the potentially very complicated nature of the ``null-related singularities'' is taken care of by the wave front set techniques. 

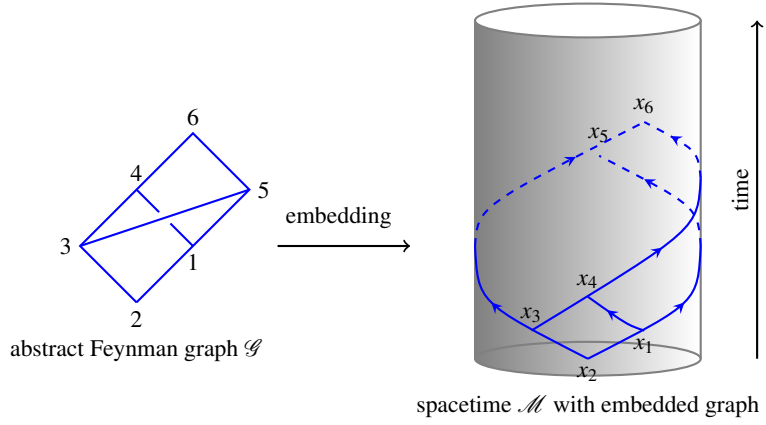
\begin{figure}
\begin{center}
\begin{tikzpicture}[scale=.75, transform shape]
\shade[left color=gray] (-2,-3) -- (2,-3) -- (2,3) -- (-2,3);
\shade[left color=gray] (0,-3) ellipse (2 and .3);
\path[fill=white] (0,3) ellipse (2 and .3);
\draw[thick,gray] (-2,-3) -- (-2,3);
\draw[thick,gray] (2,3) -- (2,-3);
\draw[thick,gray] (0,-3) ellipse (2 and .3);
\draw[thick,gray] (0,3) ellipse (2 and .3);
 
\draw[->, thick] (-5.5,-1)  -- (-3.15,-1);
\draw[->, thick] (3,-3) -- node[above,sloped]{time} (3,3);
\draw[black] (0,-3) node[below]{$x_2$};
\draw[black] (-1,-2.5) node[above]{$x_3$};
\draw[black] (0,-1.9) node[above]{$x_4$};
\draw[black] (1,-2.5) node[below]{$x_1$};
\draw[black] (1,1.2) node[above]{$x_6$};
\draw[black] (.2,.7) node[above]{$x_5$};
\draw[black] (-5.5,-.5) node[right]{embedding};
\draw[black] (0,-3.5) node[below]{spacetime $\M$ with embedded graph};
\draw[black] (-8,-2.5) node[below]{abstract Feynman graph $\eG$};
\draw[black] (-8,-2) node[below]{$2$};
\draw[black] (-9,-1) node[left]{$3$};
\draw[black] (-8,-0) node[above]{$4$};
\draw[black] (-7,1) node[above]{$6$};
\draw[black] (-6,0) node[right]{$5$};
\draw[black] (-7,-1) node[below]{$1$};

\draw[thick,blue,directed] (0,-3) .. controls (1.95,-2.0) and (1.95,-2.0) .. (2,-1);
\draw[thick,blue,directed] (0,-3) .. controls (-1.95,-2.0) and (-1.95,-2) .. (-2,-1);
\draw[thick,blue,directed] (-1,-2.5) .. controls (1.95,-.7) and (1.95,-.7) .. (2,0.2);
\draw[thick,blue,directed,dashed] (2,.2) .. controls (1.95,.7) and (1.95,.7) .. (1,1.2);
\draw[thick,blue,directed,dashed] (2,-1) .. controls (1.95,-.3) and (1.95,-.3) .. (.2,.6);
\draw[thick,blue,directed,dashed] (-2,-1) .. controls (-1.95,-.3) and (-1.95,-.3) .. (1,1.2);
\draw[thick,blue,directed] (1,-2.5) .. controls (.7,-2.4) and (.7,-2.4) .. (0,-1.9);

\draw[thick,blue] (-9,-1) -- (-6,0);
\draw[thick,blue] (-8,-2) -- (-9,-1) -- (-7,1)  -- (-6,0) -- (-8,-2);
\draw[thick,blue] (-7,-1) -- (-7.4,-.6);
\draw[thick,blue] (-7.6,-.4) -- (-8,0);

\end{tikzpicture}
\end{center}
\caption{
\label{figfeyn}
Shown here is the wave-front set of the time-ordered products~\eqref{msc} and its 
relationship with embedded Feynman graphs $\eG$ in $\M$. Through each line $e$ flows a 
`momentum' \textcolor{blue}{$p_e$} indicated by \textcolor{blue}{$\rightarrow$}, which is a parallel transported, cotangent null vector. 
At each vertex $x_i$ the corresponding vector $k_i \in T^*_{x_i} \M$ in the wave front set is characterized by 
the `momentum conservation rule' $k_i = \sum_{\rm in} \textcolor{blue}{p_e} - \sum_{\rm out} \textcolor{blue}{p_e}$
counting the momenta associated with the incoming vs. outgoing edges $e$ with opposite sign.
}
\end{figure}

An important aspect of the precise analysis~\cite{bf2,hw2} is that the wave front set of the extension can be controlled, including at the diagonals; it is characterized by the following ``{\bf microlocal
spectrum property}'': For Hadamard states $\omega$, we have 
\ben
\label{msc}
\begin{split}
& \WF\Big(\omega(\T_n( \otimes_j \phi^{k_j} )) \Big) \subset \Big\{ (x_1, k_1; ... ; x_n, k_n) \in T^* \M^n \setminus 0 \Big| \\
&  k_i = 
\sum_{e \in \eG: s(e) = i} p_e - \sum_{e \in \eG: t(e) = i} p_e \ ,  \ \ \ \ p_e \in V^\mp \ {\rm if} \ x_{s(e)/t(e)} \in J^+(x_{t(e)/s(e)}) \Big\} \ . 
\end{split}
\een
Here, one is considering embeddings of the graph $\eG$ into $\M$ such that its 
edges $e=(ij)$ are associated with null-geodesics. Their cotangent null vectors are called
$p_e$. These are future/past oriented depending on whether the edge $e=(ij)$ (oriented 
so that $s(e):=i<j=:t(e)$) is future or past directed. An illustration is given in 
fig.~\ref{figfeyn}. 

\end{document}